\documentclass{article}

\usepackage{tracecodec_preprint,times}

\usepackage[utf8]{inputenc}
\usepackage[T1]{fontenc}
\usepackage{url}
\usepackage{xurl}
\usepackage{etoolbox}
\AtBeginEnvironment{verbatim}{\fontsize{8}{9.5}\selectfont}
\usepackage{booktabs}
\usepackage{array}
\usepackage{amsfonts}
\usepackage{nicefrac}
\usepackage{microtype}
\usepackage{xcolor}
\usepackage{xspace}
\usepackage{float}
\usepackage{amsmath}
\usepackage{tikz}
\usetikzlibrary{positioning,arrows.meta,fit,calc,decorations.pathreplacing}
\usepackage{graphicx}
\usepackage{enumitem}
\usepackage{tabularx}
\usepackage[hidelinks]{hyperref}
\newcolumntype{Y}{>{\raggedright\arraybackslash}X}

\newif\ifdraft
\draftfalse
\ifdraft
  \DeclareRobustCommand{\todo}[1]{\textcolor{red}{\textbf{[TODO: #1]}}}
  
\else
  \DeclareRobustCommand{\todo}[1]{}
  
\fi

\newcommand{\reprename}{\textsc{TraceCodec}\xspace}
\newcommand{\pcap}{PCAP\xspace}
\newcommand{\flowtable}{FlowTable\xspace}

\newcommand{\action}{\mathbf{a}}
\newcommand{\latent}{\mathbf{z}}
\newcommand{\state}{\mathbf{s}}
\newcommand{\compiler}{\mathcal{C}}
\newcommand{\real}{\mathbf{p}}

\title{\reprename: A Compiler-Backed Neural Codec for Network Traffic Traces}

\author{%
Junhui Ding\textsuperscript{1} \quad
Xinchen Zhang\textsuperscript{2} \quad
Xiaohui Xie\textsuperscript{3} \quad
Shinan Liu\textsuperscript{1} \\
\textnormal{\textsuperscript{1}University of Hong Kong \qquad
\textsuperscript{2}Northeastern University \qquad
\textsuperscript{3}Tsinghua University}
}

\hypersetup{
  pdftitle={TraceCodec: A Compiler-Backed Neural Codec for Network Traffic Traces},
  pdfauthor={Junhui Ding, Xinchen Zhang, Xiaohui Xie, Shinan Liu},
  pdfsubject={Compiler-backed neural codec for stateful multi-flow network traffic traces}
}
\begin{document}

\maketitle

\begin{abstract}
Critical networking workflows require high-fidelity packet captures (PCAPs) for testing, security analysis, and protocol validation, not just statistical flow-level summaries. Recent packet generators have demonstrated protocol-constrained PCAP synthesis, but many operate on raw or near-raw packet representations. These representations can entangle learned behavioral choices with deterministic protocol rules and make packet correctness depend on post-hoc repair. We identify the decode interface as a key bottleneck and present \reprename, a compiler-backed neural codec for stateful multi-flow traces. \reprename{} lifts each parsed packet into a timed packet action with explicit flow slots and transport cues, learns a continuous per-packet latent, and uses a state-aware deterministic compiler to decode actions to PCAPs while managing endpoint assignment, TCP sequence and acknowledgment state, declared compatibility constraints, and packet rendering. On CICIDS2017 Monday, \reprename{} matches packet count, protocol composition, and flow population to within 0.03\%; raw-field baselines produce substantially larger flow-count and TCP-event errors under the same non-repair policy. In a matched flow-matching continuation setup, \reprename{} yields complete decoded suffixes in 100\% of CICIDS2017 cases and 98.44\% of MAWI cases, versus at most 10.94\% for raw-field latent controls. Structural diagnostics further show that \reprename{} preserves TCP transitions and multi-flow interleaving. Together, these results support \reprename{} as a generator-facing latent interface with a deterministic path back to usable packet traces.
\end{abstract}

\section{Introduction}
\input{figures/tracecodec_teaser}

Synthetic network traffic matters because real network traces can be very hard to share due to privacy or governance restrictions, or costly to collect~\citep{Jiang2023NetDiffusion,Chu2025NetSSM,zhou2026prvtel}. Synthetic generation can also support network evaluation and improve coverage of diverse, rare, or adversarial cases~\citep{cuppers2024flowchronicle,jin2025robustifying,gupta2025generative}. Yet useful synthetic traffic is not merely a statistically plausible table of flow summaries. Many workflows consume packet-level artifacts: traces are inspected and processed with tools such as Wireshark, DPDK, and Scapy~\citep{sikos2020packet,dpdk2026open,scapy2026manipulate}, transformed into downstream features~\citep{Holland2020nPrint,liu2023amir}, replayed in timing-sensitive environments~\citep{tcpreplay}, and used to test systems under protocol constraints. In these settings, PCAP is the object that downstream systems actually consume.

Recent work has established that packet-level generation is a realistic target. \citet{Jiang2023NetDiffusion} introduce NetDiffusion for protocol-constrained packet-trace generation, and \citet{Chu2025NetSSM} introduce NetSSM for longer, multi-flow, state-aware generation. In parallel, pretrained traffic representation models such as ET-BERT, TrafficFormer, netFound, and TrafficLLM show that useful traffic semantics can be learned for downstream analysis and cross-task generalization~\citep{Lin2022ETBERT,Zhou2025TrafficFormer,Beltiukov2026netFound,Cui2025TrafficLLM}. These advances leave open a different question: \textit{once packet traces are treated as generative targets, what representation should a packet-trace generator operate on?}

The core problem is the decode interface. Raw packet representations mix behavioral choices with fields whose values depend on flow history and packet-rendering rules. Asking one learned decoder to predict both can produce internally inconsistent packets and make final realization depend on post-hoc repair~\citep{Chu2025NetSSM,Jiang2023NetDiffusion}. A generator-facing codec therefore needs a clear contract between learned packet behavior and state-dependent realization.

More challenges arise in three properties that make packet traces usable. \textbf{(1)~Timing.} Inter-arrival times determine bursts, throughput, and latency distributions. Existing systems handle timing differently: NetSSM assigns timestamps post hoc from a ground-truth distribution, whereas netFound embeds inter-arrival and burst metadata for representation learning~\citep{Chu2025NetSSM,Beltiukov2026netFound}. A generator-facing codec instead needs an explicit timing contract. \textbf{(2)~ Network protocol statefulness.} TCP packets are constrained by a state machine: flags, sequence numbers, acknowledgments, and payload lengths must agree with the current flow state. Existing generators either learn these dependencies within the generator or apply post-generation correction, rather than exposing an explicit state-aware decode contract~\citep{Jiang2023NetDiffusion,Chu2025NetSSM}. \textbf{(3)~Multiple concurrent flows.} A capture may contain many interleaved flows, so a generator must preserve flow identity across the packet sequence. NetSSM's multi-flow evaluation, for example, reports three to five typically active flows in its Netflix workload~\citep{Chu2025NetSSM}.

Figure~\ref{fig:tracecodec_teaser} previews our approach. We introduce \reprename, a compiler-backed neural codec that uses \emph{PacketActions} rather than raw packet fields as its decode interface. Each PacketAction records the behavioral choices represented at one packet step. \textit{For timing}, we pair every action with an inter-packet time delta. \textit{For transport-state progression}, actions retain direction, TCP control behavior, payload length, and selected progression cues, while a state-aware deterministic compiler maintains endpoint assignment and sequence and acknowledgment state. \textit{For multi-flow identity}, actions refer to reusable flow slots, and the compiler maintains the state associated with each active slot. The neural codec itself remains packet-aligned, mapping each timed PacketAction to one continuous latent vector.

This separation gives downstream generative models a continuous per-packet latent space and a deterministic path back to parseable, timed, stateful PCAPs. Across decoded-PCAP evaluations, \reprename{} stays close to the reference on packet and timing statistics, while raw-field baselines incur much larger flow-count and TCP-event errors. Structural diagnostics further show that it preserves TCP transitions and multi-flow structure.

\paragraph{Contributions.}
This paper makes the following contributions:
\begin{itemize}[leftmargin=*]
    \item We identify the codec-interface problem for packet-trace generation and argue that generators should disentangle behavioral choices from deterministic protocol constraints.

    \item We introduce \reprename{}, a compiler-backed neural codec built around timed PacketActions and reusable flow slots. A packet-aligned VAE maps each action and time delta to a continuous latent, while a state-aware deterministic compiler realizes decoded actions under explicit flow and transport state.

    \item We evaluate \reprename{} as a packet-trace codec on three public datasets. After rendering reconstructed outputs to PCAP, it achieves higher packet-, timing-, transport-, and flow-level fidelity than the evaluated raw-field baselines. Under a matched downstream generator, its latent sequences also produce complete PCAP continuations substantially more often than raw-field latent controls.
\end{itemize}

\section{Background and Related Work}
\label{sec:related_work}

\paragraph{Traffic representations.}
Traffic representations span byte-level and segment-level encoders for classification and protocol understanding~\citep{Lotfollahi2020DeepPacket,Wang2017EncryptedCNN,Liu2019FSNet,Xiao2022EBSNN} and structured packet, flow, and graph views~\citep{Holland2020nPrint,Meng2022PacRep,VanEde2020FlowPrint,Huoh2023FlowGNN}. Recent systems pretrain representations over payload, datagram, flow, and cross-layer inputs~\citep{He2021PERT,Lin2022ETBERT,Zhao2023YaTC,Peng2024PTU,Wang2024NetMamba}. PACC studies protocol-aware compression~\citep{Guo2026PACC}, while TrafficFormer, netFound, and TrafficLLM pursue broader transfer and general-purpose traffic representations~\citep{Zhou2025TrafficFormer,Beltiukov2026netFound,Cui2025TrafficLLM}. These models target analysis and transfer rather than a generator-facing codec interface whose decoder must materialize timed PCAPs while maintaining flow identity and protocol state.

\paragraph{Packet generators.}
Packet generators instead synthesize network artifacts. NetShare generates IP-header traces~\citep{Yin2022NetShare}; NetDiffusion synthesizes packet-like representations with protocol-constrained realization~\citep{Jiang2023NetDiffusion}; and NetSSM models longer, multi-flow raw-packet sequences while leaving timestamps outside the learned sequence~\citep{Chu2025NetSSM}. These systems establish native packet generation, but their primary contract is an end-to-end generator rather than a reusable paired codec. \reprename addresses this complementary interface problem.

\paragraph{Codec interfaces.}
Learned codecs make high-dimensional generation tractable in images and audio~\citep{vqvae2017,rombach2022ldm,soundstream2021}. Packet traces impose an additional requirement: decoding must materialize timed packets under evolving protocol and multi-flow state. \reprename brings the codec abstraction to this setting through PacketActions and deterministic state realization. Table~\ref{tab:codec_comparison_main} summarizes the resulting interface distinction.

\begin{table*}[t]
\centering
\caption{Positioning of \reprename{} relative to traffic representations and packet generators.}
\label{tab:codec_comparison_main}
\scriptsize
\setlength{\tabcolsep}{2.8pt}
\renewcommand{\arraystretch}{1.02}
\begin{tabularx}{\textwidth}{@{}
                >{\raggedright\arraybackslash}p{0.145\textwidth}
                >{\raggedright\arraybackslash}p{0.15\textwidth}
                >{\raggedright\arraybackslash}p{0.11\textwidth}
                >{\raggedright\arraybackslash}p{0.145\textwidth}
                >{\raggedright\arraybackslash}p{0.19\textwidth}
                Y@{}}
\toprule
\textbf{Work} &
\textbf{Learned object} &
\textbf{PCAP decode} &
\textbf{Explicit timing} &
\textbf{State / multi-flow} &
\textbf{Realization contract} \\
\midrule
\multicolumn{6}{@{}l}{\textcolor{black!70}{\textsc{Traffic Representations}}} \\
\addlinespace[1pt]
~ET-BERT
~\citep{Lin2022ETBERT}
& datagram tokens
& not required
& not required
& not a decode state
& none \\
~TrafficFormer
~\citep{Zhou2025TrafficFormer}
& traffic tokens
& not required
& order, not explicit timing
& not a decode state
& none \\
~netFound
~\citep{Beltiukov2026netFound}
& traffic representation
& not required
& not required
& not a decode state
& none \\
~TrafficLLM
~\citep{Cui2025TrafficLLM}
& field/value tokens
& not required
& not required
& not a decode state
& none \\
\addlinespace[2pt]
\multicolumn{6}{@{}l}{\textcolor{black!70}{\textsc{Packet Generators}}} \\
\addlinespace[1pt]
~NetDiffusion
~\citep{Jiang2023NetDiffusion}
& packet-like image
& yes, native
& not directly generated
& single-flow traces
& protocol-constrained post-processing \\
~NetSSM
~\citep{Chu2025NetSSM}
& raw packet sequence
& yes, native
& post hoc; reference-derived
& implicit in generator state
& direct generation \\
\addlinespace[2pt]
\multicolumn{6}{@{}l}{\textcolor{black!70}{\textsc{Packet-Trace Codec}}} \\
\addlinespace[1pt]
~\textbf{TraceCodec (ours)}
& per-packet latent over timed PacketActions
& \textbf{yes}
& \textbf{yes}
& \textbf{flow slots; state in compiler}
& \textbf{deterministic compiler under explicit state} \\
\bottomrule
\end{tabularx}
\end{table*}

\section{\reprename Design}
\label{sec:design}

\subsection{Overview}
\label{subsec:overview}
\reprename separates packet behavior from its state-dependent realization. First, state-aware lifting maps each parsed packet and the current lifting state to a PacketAction and an inter-packet time delta, then advances the lifting state. Second, a packet-aligned VAE maps each timed action to one continuous latent vector. The VAE has no recurrent state or attention across packet steps. Third, a state-aware deterministic compiler executes reconstructed actions against explicit multi-flow decode state and emits concrete packets.

Figure~\ref{fig:tracecodec_overview} shows this division of responsibility. Lifting converts concrete packet fields and capture-side flow state into timed PacketActions. Flow slots preserve interaction identity in the resulting action sequence. During decoding, the compiler binds those slots to concrete flows and advances their protocol state. Downstream generators operate only on the latent sequence between these two stateful transformations.

\begin{figure*}[t]
    \centering
    \fontsize{7}{7.4}\selectfont
    \resizebox{\textwidth}{!}{%
    \begin{tikzpicture}[
        line width=0.72pt,
        line cap=round,
        line join=round,
        x=1cm,
        y=1cm,
        every node/.style={font=\sffamily\fontsize{7}{7.4}\selectfont},
        title/.style={font=\sffamily\fontsize{7}{7.4}\selectfont\bfseries, text=black},
        bandtitle/.style={font=\sffamily\fontsize{9}{9.6}\selectfont\bfseries, text=black},
        micro/.style={font=\sffamily\fontsize{5}{5.4}\selectfont, text=black},
        note/.style={font=\sffamily\fontsize{5}{5.4}\selectfont, align=center, text=black},
        pkt/.style={draw=black, fill=white, rounded corners=0.6mm, minimum width=0.25cm, minimum height=0.17cm, inner sep=0pt},
        action/.style={draw=black, fill=FlowGold!30, rounded corners=1.3mm, align=center, inner sep=2.4pt, minimum width=0.90cm, minimum height=0.62cm},
        latent/.style={circle, draw=black, fill=FlowBlue!30, minimum size=5.4mm, inner sep=0pt, font=\sffamily\fontsize{5}{5.4}\selectfont},
        chip/.style={draw=black, fill=FlowBlue!20, rounded corners=1.1mm, align=center, inner sep=2pt, minimum height=0.34cm},
        greenchip/.style={draw=black, fill=FlowGreen!25, rounded corners=1.1mm, align=center, inner sep=2pt, minimum height=0.34cm},
        arrow/.style={-{Latex[length=1.8mm,width=1.3mm]}, line width=0.63pt, draw=black},
        softarrow/.style={-{Latex[length=1.5mm,width=1.1mm]}, line width=0.60pt, draw=black, shorten >=2pt, shorten <=2pt},
        ribbonBlue/.style={line width=3.00pt, draw=black!8, line cap=round},
        ribbonGreen/.style={line width=3.00pt, draw=black!8, line cap=round},
        railA/.style={line width=0.98pt, draw=black},
        railB/.style={line width=0.98pt, draw=black},
        railC/.style={line width=0.98pt, draw=black}
    ]

    \definecolor{FlowBlue}{HTML}{5B8FB9}
    \definecolor{FlowGold}{HTML}{EDB750}
    \definecolor{FlowViolet}{HTML}{B294CE}
    \definecolor{FlowGreen}{HTML}{83B777}
    \definecolor{AccentBlue}{HTML}{2385D1}
    \definecolor{SuccessGreen}{HTML}{27813F}
    \definecolor{HeaderGray}{HTML}{ECECEC}
    \definecolor{HeaderBlue}{HTML}{D3E5F5}
    \definecolor{HeaderGreen}{HTML}{DDEBDD}

    \filldraw[fill=white, draw=black, rounded corners=3.0mm]
        (-0.10,3.62) rectangle (17.30,7.06);
    \filldraw[fill=white, draw=black, rounded corners=3.0mm]
        (-0.10,-0.02) rectangle (17.30,3.40);
    \begin{scope}
        \clip[rounded corners=3.0mm] (-0.10,3.62) rectangle (17.30,7.06);
        \fill[HeaderBlue] (-0.10,6.40) rectangle (17.30,7.06);
    \end{scope}
    \begin{scope}
        \clip[rounded corners=3.0mm] (-0.10,-0.02) rectangle (17.30,3.40);
        \fill[HeaderGreen] (-0.10,2.80) rectangle (17.30,3.40);
    \end{scope}
    \draw[draw=black, line width=0.90pt, rounded corners=3.0mm]
        (-0.10,3.62) rectangle (17.30,7.06);
    \draw[draw=black, line width=0.90pt, rounded corners=3.0mm]
        (-0.10,-0.02) rectangle (17.30,3.40);
    \draw[draw=black, line width=0.75pt] (-0.10,6.40) -- (17.30,6.40);
    \draw[draw=black, line width=0.75pt] (-0.10,2.80) -- (17.30,2.80);
    \node[bandtitle, anchor=west] at (0.28,6.74) {Encode: Packet Trace \(\rightarrow\) Timed Packet Actions \(\rightarrow\) Latents};
    \node[bandtitle, anchor=west] at (0.28,3.14) {Decode: Latents \(\rightarrow\) Timed Packet Actions \(\rightarrow\) Protocol-Constrained Trace};

    \draw[ribbonBlue] (1.15,5.18) .. controls (3.65,5.52) and (5.35,4.70) .. (8.95,5.18)
        .. controls (10.65,5.62) and (11.55,4.74) .. (13.10,5.18)
        .. controls (14.02,5.40) and (15.15,5.20) .. (16.55,5.18);
    \draw[ribbonGreen] (1.20,1.42) .. controls (2.58,1.72) and (3.72,1.26) .. (5.10,1.42)
        .. controls (6.65,1.60) and (7.55,1.18) .. (9.02,1.42)
        .. controls (10.98,1.72) and (12.90,1.14) .. (16.18,1.42);

    \node[title] at (1.62,3.93) {Multi-Flow \pcap};
    \draw[railA] (0.70,5.50) -- (2.55,5.50);
    \draw[railB] (0.86,5.14) -- (2.38,5.14);
    \draw[railC] (1.02,4.78) -- (2.48,4.78);
    \foreach \x in {0.86,1.20,1.62,2.08,2.40} \node[pkt, fill=FlowBlue] at (\x,5.50) {};
    \foreach \x in {1.04,1.48,1.94,2.26} \node[pkt, fill=FlowGold] at (\x,5.14) {};
    \foreach \x in {1.20,1.72,2.22} \node[pkt, fill=FlowViolet] at (\x,4.78) {};

    \draw[draw=black, fill=white, rounded corners=1.8mm, line width=0.60pt]
        (3.18,4.48) rectangle (6.06,5.90);
    \draw[draw=black] (4.72,4.66) -- (4.72,5.72);
    \node[micro, text=black] at (3.88,5.70) {Slot Map};
    \node[micro, text=black] at (5.30,5.70) {State};
    \foreach \y/\col/\slot in {
        5.42/blue!62!black/\(F7\),
        5.14/green!45!black/\(F2\),
        4.86/violet!55!black/\(F9\)
    } {
        \draw[draw=black, line width=0.99pt] (3.38,\y) -- (3.80,\y);
        \node[pkt, draw=black, fill=white, minimum width=0.18cm, minimum height=0.13cm] at (3.59,\y) {};
        \draw[softarrow, draw=black] (3.84,\y) -- (4.06,\y);
        \node[draw=black, fill=AccentBlue, text=white, rounded corners=0.9mm, inner sep=1.3pt, font=\sffamily\tiny, minimum width=0.50cm] at (4.33,\y) {\slot};
    }
    \node[draw=black, fill=HeaderGray, rounded corners=1.0mm, align=center, inner sep=2.0pt, font=\sffamily\fontsize{5}{5.4}\selectfont, minimum width=1.12cm, minimum height=0.78cm] at (5.42,5.10)
        {\(S_t\)\\[1pt]{\fontsize{5.5}{6.2}\selectfont active flows}\\[1pt]{\fontsize{5.5}{6.2}\selectfont protocol state}};
    \draw[softarrow] (2.66,5.18) -- (3.28,5.18);

    \node[title] at (8.63,3.93) {Packet Actions};
    \foreach \x/\dt/\slot/\ev in {
        7.13/.03/F7/SYN,
        8.33/.11/F2/DATA,
        9.53/.02/F7/ACK,
        10.73/.08/F9/FIN
    } {
        \node[action, minimum width=1.16cm] at (\x,5.28)
            {\(\Delta t\) \dt\\[-1pt]\slot\\[-1pt]\texttt{\ev}};
    }
    \draw[draw=black, line width=0.83pt] (6.60,4.45) -- (11.26,4.45);
    \foreach \x in {7.13,8.33,9.53,10.73}
        \draw[draw=black, line width=0.68pt] (\x,4.35) -- (\x,4.55);
    \draw[softarrow] (5.96,5.18) -- (6.54,5.18);

    \node[title] at (12.79,3.93) {Neural Encoder};
    \draw[draw=black, fill=white, rounded corners=2mm, line width=0.83pt]
        (11.83,4.45) -- (13.27,4.82) -- (13.27,5.54) -- (11.83,5.91) -- cycle;
    \draw[draw=black, line width=0.60pt] (12.09,4.70) -- (13.01,4.95);
    \draw[draw=black, line width=0.60pt] (12.09,5.66) -- (13.01,5.41);
    \node[chip, minimum width=0.62cm] at (12.23,5.40) {\(a_t\)};
    \node[chip, minimum width=0.68cm] at (12.23,5.02) {\(\delta_t\)};
    \draw[softarrow, shorten >=1pt, shorten <=1pt] (11.26,5.18) -- (11.79,5.18);

    \node[title, align=center] at (15.18,3.93) {Latents};
    \draw[draw=black, line width=1.20pt] (14.18,5.18) -- (16.10,5.18);
    \node[latent] at (14.18,5.18) {\(\latent_1\)};
    \node[latent] at (14.82,5.18) {\(\latent_2\)};
    \node[latent] at (15.46,5.18) {\(\latent_3\)};
    \node[latent] at (16.10,5.18) {\(\latent_4\)};
    \draw[softarrow] (13.25,5.18) -- (13.82,5.18);

    \node[title] at (1.68,0.29) {Latents};
    \draw[draw=black, line width=1.20pt] (0.72,1.42) -- (2.66,1.42);
    \node[latent] at (0.72,1.42) {\(\latent_1\)};
    \node[latent] at (1.36,1.42) {\(\latent_2\)};
    \node[latent] at (2.00,1.42) {\(\latent_3\)};
    \node[latent] at (2.64,1.42) {\(\latent_4\)};

    \node[title] at (4.56,0.29) {Neural Decoder};
    \draw[draw=black, fill=white, rounded corners=2mm, line width=0.83pt]
        (3.56,0.70) -- (4.06,2.06) -- (5.28,1.86) -- (5.28,0.98) -- cycle;
    \node[latent, minimum size=5.0mm] (zfan) at (3.84,1.42) {\(\latent_t\)};
    \node[chip, minimum width=1.14cm] (actionfields) at (5.02,1.72) {PacketAction\\[-1pt]{\fontsize{5}{5.4}\selectfont fields}};
    \node[chip, minimum width=0.84cm] (time) at (5.02,1.10) {Timing\\[-1pt]{\fontsize{5}{5.4}\selectfont \(\Delta t\)}};
    \draw[softarrow] (zfan) -- (actionfields.west);
    \draw[softarrow] (zfan) -- (time.west);
    \draw[softarrow] (2.92,1.42) -- (3.44,1.42);

    \node[title] at (7.67,0.29) {Decoded Actions};
    \foreach \x/\ev in {6.67/SYN,7.37/DATA,8.07/ACK,8.77/FIN} {
        \node[action, minimum width=0.66cm, minimum height=0.56cm, inner xsep=1.6pt] at (\x,1.42)
            {\(\widehat{\Delta t}\)\\[-1pt]slot\\[-1pt]\texttt{\ev}};
    }
    \draw[softarrow] (5.58,1.42) -- (6.27,1.42);

    \begin{scope}[yshift=0.36cm]
    \node[title] at (11.40,-0.07) {Deterministic State Realization};
    \draw[draw=black, fill=white, rounded corners=2.2mm, line width=0.83pt]
        (9.78,0.38) rectangle (13.24,2.20);
    \node[micro] at (10.42,1.96) {State Store};
    \node[draw=black, fill=white, rounded corners=1.0mm, line width=0.68pt, minimum width=1.06cm, minimum height=0.72cm, inner sep=0pt] (statestore) at (10.42,1.40) {};
    \draw[draw=black, line width=0.60pt]
        (9.95,1.52) -- (10.89,1.52);
    \draw[draw=black, line width=0.60pt]
        (9.95,1.28) -- (10.89,1.28);
    \node[micro] at (10.42,1.64) {slot};
    \node[micro] at (10.42,1.40) {template};
    \node[micro] at (10.42,1.16) {seq/ack};

    \node[draw=black, fill=white, rounded corners=1.0mm, minimum width=1.56cm, minimum height=0.34cm, inner sep=0pt] (transitionbar) at (11.42,0.70) {};
    \draw[draw=black, line width=0.68pt] (11.19,0.53) -- (11.19,0.87);
    \draw[draw=black, line width=0.68pt] (11.65,0.53) -- (11.65,0.87);
    \node[micro] at (10.94,0.70) {SYN};
    \node[micro] at (11.42,0.70) {EST};
    \node[micro] at (11.90,0.70) {DATA};
    \draw[softarrow, draw=SuccessGreen] (statestore.south) -- (10.42,0.70) -- (transitionbar.west);

    \node[micro] at (12.56,1.96) {Packet Fields};
    \node[draw=black, fill=white, rounded corners=1.0mm, line width=0.68pt, minimum width=0.78cm, minimum height=0.76cm, inner sep=0pt] (packetfields) at (12.56,1.34) {};
    \draw[draw=black, line width=0.60pt]
        (12.22,1.47) -- (12.90,1.47);
    \draw[draw=black, line width=0.60pt]
        (12.22,1.22) -- (12.90,1.22);
    \node[micro] at (12.56,1.59) {IP};
    \node[micro] at (12.56,1.34) {TCP};
    \node[micro] at (12.56,1.09) {cksum};
    \draw[softarrow, draw=SuccessGreen] (transitionbar.east) -- (12.56,0.70) -- (packetfields.south);
    \end{scope}
    \draw[softarrow] (9.15,1.42) -- (9.70,1.42);

    \node[title] at (15.06,0.29) {Multi-Flow \pcap};
    \draw[railA, draw=SuccessGreen] (14.00,1.74) -- (16.16,1.74);
    \draw[railB, draw=SuccessGreen] (14.20,1.36) -- (15.88,1.36);
    \draw[railC, draw=SuccessGreen] (14.38,0.98) -- (16.04,0.98);
    \foreach \x in {14.22,14.66,15.10,15.56,15.96} \node[pkt, draw=SuccessGreen, fill=FlowBlue] at (\x,1.74) {};
    \foreach \x in {14.40,14.94,15.38,15.76} \node[pkt, draw=SuccessGreen, fill=FlowGold] at (\x,1.36) {};
    \foreach \x in {14.60,15.16,15.70} \node[pkt, draw=SuccessGreen, fill=FlowViolet] at (\x,0.98) {};
    \draw[softarrow, draw=SuccessGreen] (13.32,1.42) -- (13.80,1.42);

    \end{tikzpicture}%
    }
    \caption{Encoding and decoding phases of \reprename }
    \label{fig:tracecodec_overview}
\end{figure*}

\subsection{Packet Actions}
\label{subsec:action_ir}

A \emph{PacketAction} is a structured record of the behavioral choices represented at one packet step. It records the flow slot, protocol, direction, control event, payload behavior, and selected transport cues. It omits fields reconstructed from decode state, including concrete endpoints and absolute TCP sequence numbers. A PacketAction is therefore a codec IR, not a compressed packet header.

\begin{table}[t]
\centering
\caption{PacketAction schema. Rows group fields only for presentation; each packet step uses one fixed-width action record.}
\label{tab:action_schema}
\footnotesize
\setlength{\tabcolsep}{4pt}
\renewcommand{\arraystretch}{1.0}
\begin{tabularx}{\linewidth}{@{}p{0.28\linewidth}Y@{}}
\toprule
\textbf{Group} & \textbf{Fields and decode role} \\
\midrule
Flow, family, and event identity &
\texttt{flow\_token}, \texttt{flow\_evt}, \texttt{proto}, \texttt{ip\_family}, \texttt{dir}. Identify the bounded active-flow slot, protocol path, IP family, and flow lifecycle role. \\
\addlinespace[2pt]
TCP control semantics &
\texttt{tcp\_ctrl}. Represent TCP control events such as SYN, SYN-ACK, FIN, and RST without requiring raw flag prediction. \\
\addlinespace[2pt]
Transport fields &
\texttt{tcp\_opt\_profile}, \texttt{tcp\_win\_d0..d3}, \texttt{tcp\_ack\_adv\_d0..d7}, \texttt{tcp\_ack\_unseen}, \texttt{tcp\_seq\_delta\_sign}, \texttt{tcp\_seq\_delta\_d0..d7}. Encode option profile, receive-window value, bounded acknowledgment advancement, unseen-ACK state, and bounded sequence progression. \\
\addlinespace[2pt]
Packet fields &
\texttt{ttl\_res}, \texttt{icmp\_type}, \texttt{icmp\_code}, \texttt{l4\_payload\_len\_d0..d3}. Encode IP-family-specific values, ICMP control fields, and layer-4 payload length. \\
\addlinespace[2pt]
\bottomrule
\end{tabularx}
\end{table}

\paragraph{Lifting packets.}
After ordinary PCAP parsing, the lifter processes each eligible parsed packet against the current lifting state. It maps the packet's concrete flow to a reusable slot, derives the selected relative transport cues, emits a discrete PacketAction \(\action_t\) and an inter-packet time delta \(\delta_t\), and then advances the lifting state. The action and time delta form one packet-aligned codec record. Table~\ref{tab:action_schema} lists the discrete action fields.

The \texttt{flow\_token} field names a reusable slot in a bounded active-flow vocabulary, not an endpoint identifier. It preserves which active interaction the packet belongs to while leaving concrete endpoint assignment to the compiler.

The \texttt{flow\_evt} field controls the slot lifecycle. \textsc{NEW} initializes a slot for a flow, \textsc{CONTINUE} refers to the active flow in that slot, and \textsc{END} releases the slot after its final packet. A generation index distinguishes successive flows that reuse the same slot.

\paragraph{Behavioral variables.}
The IR retains variables whose changes affect the decoded trace. Packet order is preserved by the action sequence, timing by \(\delta_t\), and flow identity by reusable slots. A decoded PacketAction does not determine every field of the realized packet. Concrete endpoints and absolute TCP sequence and acknowledgment numbers depend on decode state; derived lengths and checksums are recomputed during rendering. Application payload semantics remain outside the codec.

\paragraph{State-aware compilation.}
A PacketAction is therefore a residual rather than a complete packet: the packet-aligned VAE reconstructs each timed action independently, while the compiler combines it with the current decode state to emit a packet and advance that state. Section~\ref{subsec:compiler} formalizes this transition.

\subsection{Packet Compiler}
\label{subsec:compiler}

The compiler executes one decoded action at a time. The global decode state \(\state_t\) contains the active flow slots and their protocol state. Given this state, the compiler emits one packet and advances the state:
\[
(\hat{\real}_t, \state_{t+1})
=
\compiler(\state_t, \hat{\action}_t, \hat{\delta}_t).
\]
Starting from an empty state, repeated application emits the reconstructed trace within the codec scope. The compiler realizes state-dependent consequences through FlowTable lookup, transport-state updates, and packet rendering; it is part of the decode contract, not a repair stage applied after generation.

\paragraph{Flow templates.}
The compiler first resolves the slot encoded by \texttt{flow\_token}. A deterministic \flowtable binds the slot and its generation index to a synthetic endpoint template. The decoded direction selects the endpoint orientation. Reusable slots give the interface a fixed active-flow capacity without making the VAE reconstruct raw five-tuples.

\paragraph{Transport state.}
For TCP traffic, the compiler maintains per-flow, per-direction transport state. It assigns initial sequence numbers, tracks sequence and acknowledgment progress, and updates the state after each decoded control or data event. The PacketAction supplies the event and its bounded progression cues; the compiler performs the corresponding arithmetic.

UDP and ICMP require less state, but their realization remains explicit. The compiler instantiates endpoints and applies the decoded direction, payload length, and ICMP control fields. In every protocol path, the VAE predicts bounded behavior and the compiler constructs the packet under explicit state.

\paragraph{Constrained rendering.}
Before rendering TCP packets, the compiler applies a fixed set of local compatibility constraints. It clears payload length on control events and acknowledgment advancement on a pure SYN. These rules resolve only the declared conflicts inside an action. They do not correct an incorrect protocol, flow event, slot, direction, or TCP control event.

After flow and transport interpretation, the compiler renders a concrete packet using link, network, and transport headers. Payload bytes are deterministic synthetic material by default; their lengths are controlled by the decoded action, but the bytes themselves do not reconstruct application-level content. \reprename therefore targets parseable, transport-consistent packet traces rather than byte-identical application conversations.

\subsection{Latent Codec}
\label{subsec:model}

With the PacketAction IR and deterministic compiler fixed, the learned component is a packet-aligned VAE~\citep{Kingma2014VAE}. Each forward pass processes one PacketAction and its time delta. The VAE has no recurrent hidden state and no attention across packet steps. A trace becomes an ordered latent sequence with one vector per packet step.

\paragraph{Formulation.}
Let \(\action_t\) denote a PacketAction and \(\delta_t\) its inter-packet time delta. The encoder defines
\[
q_{\phi}(\latent_t \mid \action_t, \delta_t),
\]
and the decoder defines
\[
p_{\theta}(\action_t, \delta_t \mid \latent_t).
\]
The decoder produces \(\hat{\action}_t\) and \(\hat{\delta}_t\); protocol-state transitions remain the compiler's responsibility.

The encoder embeds the fields of one timed PacketAction as tokens, applies a compact Transformer within that record, and maps the result to Gaussian latent parameters. Field-specific classification heads reconstruct the fixed-width PacketAction, and a separate regression head reconstructs timing. Neither the encoder nor the decoder carries state across packet steps. We use a conventional backbone so that the evaluation focuses on the representation and decode interface.

\paragraph{Latent interface.}
The continuous latent layer is the interface exposed to downstream generators. It keeps timing with packet behavior and gives a sequence model one vector per packet step. Other latent families can use the same PacketAction IR because the compiler contract does not depend on the VAE architecture.

\paragraph{Training objective.}
The model is trained with a variational objective
\[
\mathcal{L}
=
\mathcal{L}_{\mathrm{recon}}
+ \lambda_{\mathrm{time}}\mathcal{L}_{\mathrm{time}}
+ \beta \mathcal{L}_{\mathrm{KL}},
\]
where \(\mathcal{L}_{\mathrm{recon}}\) aggregates field-wise reconstruction losses over the PacketAction record, \(\mathcal{L}_{\mathrm{time}}\) measures timing reconstruction error, and \(\mathcal{L}_{\mathrm{KL}}\) regularizes the latent space. Reconstruction is computed in packet-action space rather than packet space; decoded packets are obtained only after passing reconstructed packet actions through the deterministic compiler.

The KL term regularizes the generator-facing latent space; the ablation study separates this role from local reconstruction fidelity.

\section{Evaluation}
\label{sec:experiments}

\subsection{Experimental Setup}

\paragraph{Datasets and baselines.}
We evaluate \reprename on CICIDS2017 Monday~\citep{Sharafaldin2018CICIDS2017} and MAWI \texttt{202004071400}~\citep{Cho2000WIDERepository}, which together cover enterprise and backbone traffic. To broaden the evaluation to a distinct VPN/non-VPN application-traffic setting, we additionally evaluate on VNAT~\citep{Jorgensen2024VNAT}. Raw-field baselines use four published tabular generative models adapted to the same encode--decode task: TVAE~\citep{Xu2019CTGAN}, TabSyn-VAE~\citep{Zhang2024TabSyn}, GOGGLE~\citep{Liu2023GOGGLE}, and TTVAE~\citep{Wang2025TTVAE}.

\paragraph{Evaluation protocol.}
We render every raw-field baseline output under a shared non-repair policy: packet-local recomputation is allowed, but endpoint inference, sequence-level repair, TCP-state repair, and use of \reprename{}'s compiler are not. We also include an \emph{Oracle} control, which lowers ground-truth PacketActions and timing through the deterministic compiler without a learned bottleneck. All neural codecs are trained from scratch; none uses an external pretrained backbone. All decoded-PCAP metrics are computed after packet rendering. Count columns report relative errors, TV columns compare distributions, and TCP-event distance compares diagnostic-event rate profiles rather than invalid-packet rates. Dataset construction and splitting, baseline adaptation, model capacity, runtime, training budgets, and metric definitions are given in Appendices~\ref{app:setup}, \ref{app:baselines}, \ref{app:metrics}, and~\ref{app:repro}.

\subsection{PCAP Reconstruction Fidelity}

\begin{table*}[t!]
\centering
\caption{Decoded-PCAP fidelity on held-out test sets. Lower is better; bold marks the best learned method, and ``--'' denotes no valid output. Oracle is a nonlearned control.}
\label{tab:eval_main_results}
\scriptsize
\setlength{\tabcolsep}{2.6pt}
\renewcommand{\arraystretch}{1.02}
\begin{tabular*}{\textwidth}{@{\extracolsep{\fill}}l l r r r r r r@{}}
\toprule
\textbf{Dataset} &
\textbf{Method} &
\shortstack{\textbf{TCP event}\\\textbf{dist. (pp) \(\downarrow\)}} &
\shortstack{\textbf{Packet count}\\\textbf{err. (\%) \(\downarrow\)}} &
\shortstack{\textbf{Protocol TV}\\\textbf{(\%) \(\downarrow\)}} &
\shortstack{\textbf{IAT TV}\\\textbf{(\%) \(\downarrow\)}} &
\shortstack{\textbf{Flow count}\\\textbf{err. (\%) \(\downarrow\)}} &
\shortstack{\textbf{Flow duration}\\\textbf{TV (\%) \(\downarrow\)}} \\
\midrule
\shortstack[l]{CICIDS2017\\Monday} &
TVAE &
72.73 &
37.83 &
5.16 &
14.83 &
25.19 &
45.36 \\
&
TabSyn-VAE &
66.74 &
2.97 &
0.62 &
1.08 &
46.32 &
34.76 \\
&
GOGGLE &
-- &
-- &
-- &
-- &
-- &
-- \\
&
TTVAE &
53.19 &
37.65 &
5.34 &
14.40 &
104.98 &
51.03 \\
&
\textbf{\reprename} &
\textbf{14.51} &
\textbf{0.00} &
\textbf{0.00} &
\textbf{0.84} &
\textbf{0.03} &
\textbf{9.23} \\
&
Oracle &
2.17 &
0.00 &
0.00 &
0.04 &
0.00 &
0.01 \\
\midrule
\shortstack[l]{MAWI\\\texttt{202004071400}} &
TVAE &
52.67 &
14.21 &
8.06 &
29.33 &
91.31 &
29.87 \\
&
TabSyn-VAE &
53.63 &
28.05 &
6.71 &
5.92 &
76.07 &
10.73 \\
&
GOGGLE &
-- &
-- &
-- &
-- &
-- &
-- \\
&
TTVAE &
50.37 &
61.96 &
18.27 &
26.45 &
81.29 &
17.54 \\
&
\textbf{\reprename} &
\textbf{2.31} &
\textbf{0.00} &
\textbf{0.00} &
\textbf{0.38} &
\textbf{0.07} &
\textbf{0.25} \\
&
Oracle &
1.55 &
0.00 &
0.00 &
0.01 &
0.00 &
0.00 \\
\midrule
VNAT &
TVAE &
113.68 &
51.14 &
3.32 &
47.64 &
61.02 &
57.09 \\
&
TabSyn-VAE &
74.96 &
4.45 &
0.51 &
7.13 &
30.07 &
48.38 \\
&
GOGGLE &
-- &
-- &
-- &
-- &
-- &
-- \\
&
TTVAE &
122.86 &
31.38 &
2.03 &
59.14 &
384.63 &
55.51 \\
&
\textbf{\reprename} &
\textbf{0.14} &
\textbf{0.00} &
\textbf{0.00} &
\textbf{1.08} &
\textbf{8.69} &
\textbf{6.84} \\
&
Oracle &
0.00 &
0.00 &
0.00 &
0.02 &
0.00 &
0.00 \\
\bottomrule
\end{tabular*}
\end{table*}

\begin{figure*}[t]
\centering
\includegraphics[width=1.00\textwidth]{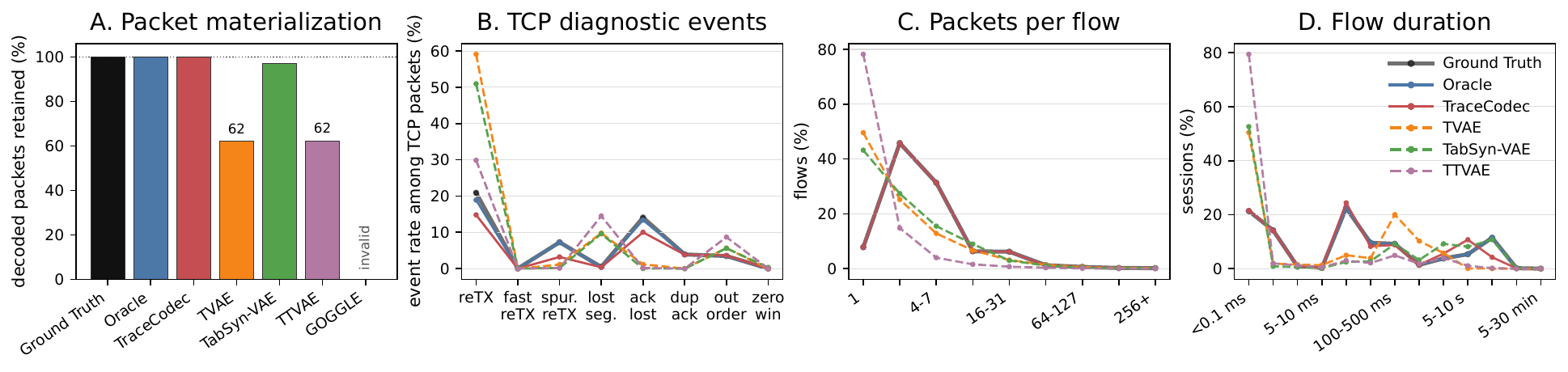}
\caption{Protocol state preservation analysis on CICIDS2017 Monday.}
\label{fig:decoded_pcap_claim_panels}
\end{figure*}

Table~\ref{tab:eval_main_results} evaluates fully rendered PCAPs. The Oracle has zero packet-count, protocol, and flow-count error on all three datasets, showing that the symbolic path retains this structure before learning. After neural reconstruction, \reprename{} has zero count and protocol error on all three datasets; its flow-count errors are 0.03\%, 0.07\%, and 8.69\% on CICIDS2017, MAWI, and VNAT. The largest remaining errors are TCP events and flow duration on CICIDS2017 and flow count on VNAT. Raw-field baselines instead lose packets, fragment flows, or distort transport behavior. Figure~\ref{fig:decoded_pcap_claim_panels} exposes these failure modes through packet materialization, TCP diagnostics, and flow distributions.

\subsection{TCP Transitions and Multi-Flow Interleaving}

\begin{figure*}[t]
\centering
\includegraphics[width=1.00\textwidth]{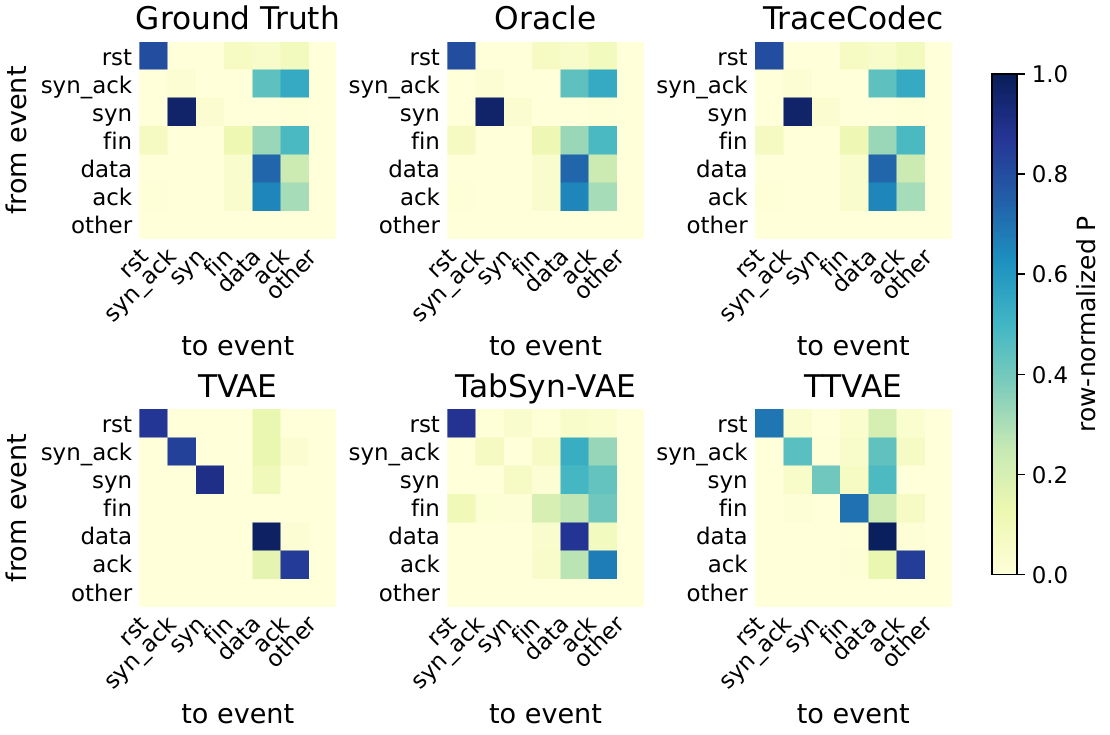}
\caption{TCP transition and state consistency on CICIDS2017 Monday. Each heatmap is a row-normalized transition matrix over coarse TCP event labels within bidirectional sessions. }
\label{fig:tcp_transition_state_consistency}
\end{figure*}

\paragraph{TCP transition consistency.}
Marginal event rates do not capture event ordering within a session. Figure~\ref{fig:tcp_transition_state_consistency} therefore measures event-to-event transitions within bidirectional sessions. The Oracle and learned \reprename{} have weighted row-wise total-variation distances of \(3.66\times 10^{-5}\) and \(2.76\times 10^{-5}\) from the reference, versus \(0.229\)--\(0.353\) for the raw-field baselines. Thus, \reprename{} closely matches the reference distribution of TCP-event transitions beyond their marginal rates.

\begin{figure*}[t]
\centering
\includegraphics[width=\textwidth]{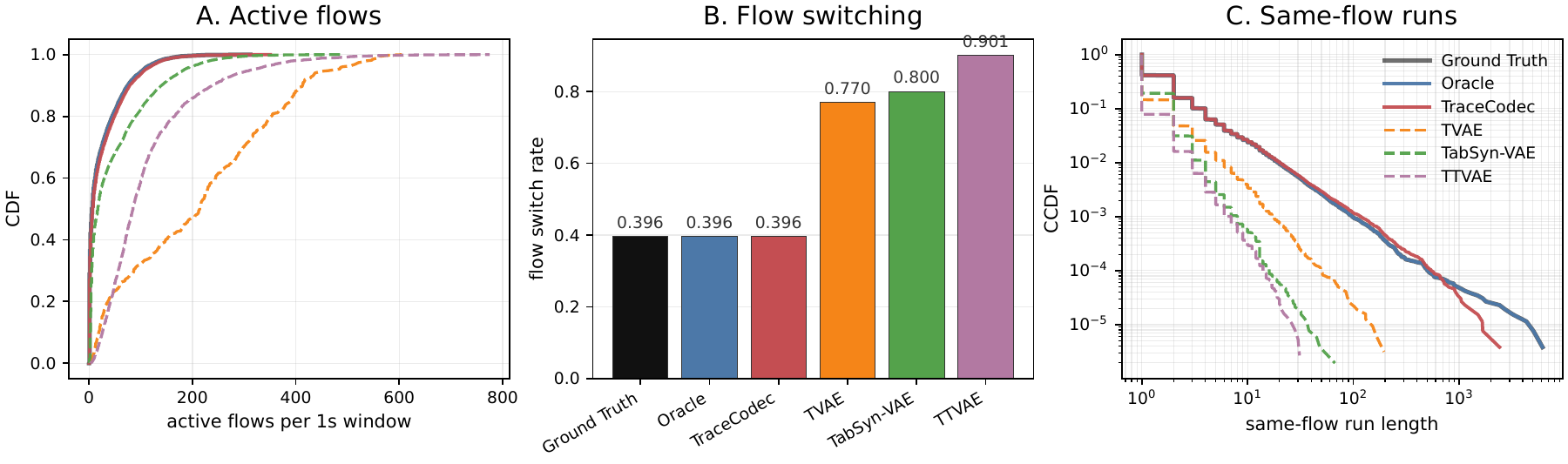}
\caption{Active-flow and multi-flow interleaving on CICIDS2017 Monday.}
\label{fig:active_flow_interleaving}
\end{figure*}

\paragraph{Multi-flow interleaving.}
Figure~\ref{fig:active_flow_interleaving} tests whether decoded packets preserve concurrent sessions without spurious fragmentation. The reference, Oracle, and learned \reprename{} have similar flow-switch rates (\(\approx 0.396\)) and same-flow run lengths (mean \(\approx 2.53\)). Raw-field baselines switch flows more often, with rates of \(0.770\)--\(0.901\), and inflate or lose sessions. \reprename{} therefore retains the reference interleaving pattern without fragmenting flows.

\subsection{Latent Interface for Traffic Continuation}

This experiment evaluates the usability of the latent decode interface, not a complete end-to-end traffic-generation system. We use the same conditional flow-matching architecture~\citep{Lipman2023FlowMatching} and protocol across representations. The models contain 135.15--135.22M parameters, a difference below 0.05\%. For each matched window, the generator observes a 1024-token prefix and produces a 1024-token suffix, which we decode and compare with the held-out suffix PCAP. We include the standalone TVAE and TabSyn-VAE latents; TTVAE requires input-derived encoder memory, and GOGGLE produces no valid rows under the shared non-repair contract.

We count a suffix as usable when its rendered PCAP contains exactly the requested number of packets; valid-packet coverage and distributional fidelity are reported separately. Coverage-adjusted total variation (CA-TV) compares each decoded distribution with its reference while charging missing or invalid packet mass. Ordinary TV can instead make a method appear accurate after it drops difficult packets.

\begin{table*}[t]
\centering
\caption{Latent-interface results on CICIDS2017 Monday over 128 matched windows. Entries are case-level medians except usable suffixes; CA denotes coverage adjustment.}
\label{tab:latent_interface_check}
\scriptsize
\setlength{\tabcolsep}{4.0pt}
\renewcommand{\arraystretch}{1.10}
\resizebox{\textwidth}{!}{%
\begin{tabular}{@{}l r r r r r r@{}}
\toprule
\textbf{Representation} &
\textbf{Usable suffixes (\%) \(\uparrow\)} &
\textbf{Valid pkt coverage (\%) \(\uparrow\)} &
\textbf{CA IAT TV (\%) \(\downarrow\)} &
\textbf{CA pkt-size TV (\%) \(\downarrow\)} &
\textbf{CA flow-pkts TV (\%) \(\downarrow\)} &
\textbf{CA sess-dur TV (\%) \(\downarrow\)} \\
\midrule
\textbf{\reprename} &
{\bfseries 100.00} &
{\bfseries 100.00} &
{\bfseries 13.00} &
{\bfseries 18.85} &
{\bfseries 39.35} &
{\bfseries 40.44} \\
TVAE &
0.00 &
63.18 &
41.07 &
49.76 &
64.80 &
57.75 \\
TabSyn-VAE &
7.03 &
98.29 &
16.09 &
25.59 &
42.99 &
42.59 \\
\bottomrule
\end{tabular}
}
\end{table*}

Table~\ref{tab:latent_interface_check} shows that \reprename{} produces usable suffixes in all 128 cases with 100\% median valid-packet coverage, compared with 0\% usable suffixes for TVAE and 7.03\% for TabSyn-VAE. It also has the lowest coverage-adjusted timing and structural distances. Continuation remains imperfect, but sampled \reprename{} latents materialize as complete PCAPs far more consistently.

\subsection{Codec and Realization Ablations}

\begin{table*}[t]
\centering
\caption{Codec and realization ablations on CICIDS2017 Monday. The upper block compares codec variants against its full-model control. The lower block compares two realizers on the same frozen codec reconstruction and belongs to a separate paired run. Boxes mark targeted effects; bold marks the better realizer; ``--'' denotes a metric not reported for that block. KL rate is in nats per packet.}
\label{tab:eval_ablations}
\scriptsize
\setlength{\tabcolsep}{2.2pt}
\renewcommand{\arraystretch}{1.14}
\newcommand{\abmark}[1]{\begingroup\setlength{\fboxsep}{1.0pt}\setlength{\fboxrule}{0.45pt}\fbox{\strut #1}\endgroup}
\begin{tabular*}{\textwidth}{@{\extracolsep{\fill}}l r r r r r r r@{}}
\toprule
\textbf{Variant / realizer} &
\shortstack{\textbf{Action field}\\\textbf{acc. (\%) \(\uparrow\)}} &
\shortstack{\textbf{TCP event}\\\textbf{dist. (pp) \(\downarrow\)}} &
\shortstack{\textbf{IAT TV}\\\textbf{(\%) \(\downarrow\)}} &
\shortstack{\textbf{Flow count}\\\textbf{err. (\%) \(\downarrow\)}} &
\shortstack{\textbf{Flow duration}\\\textbf{TV (\%) \(\downarrow\)}} &
\shortstack{\textbf{Valid pkt}\\\textbf{coverage (\%) \(\uparrow\)}} &
\shortstack{\textbf{KL rate}\\\textbf{(nats/pkt)}} \\
\midrule
\multicolumn{8}{@{}l}{\textcolor{black!70}{\textsc{Codec Components}}} \\
\addlinespace[1pt]
Full \reprename &
99.73 &
13.79 &
0.59 &
0.40 &
8.41 &
-- &
167.91 \\
\textsc{w/o} KL &
99.58 &
14.57 &
0.49 &
0.41 &
8.64 &
-- &
\abmark{5222.77} \\
\textsc{w/o} fine timing &
99.77 &
12.89 &
\abmark{13.37} &
0.40 &
5.90 &
-- &
152.41 \\
\textsc{w/o} TCP cues &
99.95 &
\abmark{138.85} &
0.51 &
0.40 &
8.10 &
-- &
94.22 \\
\addlinespace[4pt]
\multicolumn{8}{@{}l}{\textcolor{black!70}{\textsc{Deterministic vs. Learned Realization}}} \\
\addlinespace[1pt]
Deterministic compiler &
-- &
\textbf{12.66} &
\textbf{0.49} &
\textbf{0.04} &
\textbf{9.13} &
\textbf{100.00} &
-- \\
Causal LSTM realizer &
-- &
43.52 &
85.89 &
631.22 &
77.61 &
89.75 &
-- \\
\bottomrule
\end{tabular*}
\end{table*}

\paragraph{Codec components.}
The upper block of Table~\ref{tab:eval_ablations} separates timing, TCP progression and window cues, and latent regularization within a matched ablation run. Its full-model row is the control for these variants and is not the separate reconstruction artifact reported in Table~\ref{tab:eval_main_results}. Coarsening timing raises IAT TV from 0.59\% to 13.37\%. Removing TCP progression and window cues leaves action accuracy at 99.95\% but raises TCP-event distance to 138.85 percentage points. Removing KL changes reconstruction little but raises the posterior KL rate from 167.91 to 5222.77 nats per packet. Thus, action accuracy alone does not establish timing fidelity, transport behavior, or latent regularity.

\paragraph{Realization mechanism.}
Because PacketActions omit state-dependent packet fields, \reprename{} uses a deterministic compiler for realization. We compare it with a 4.27M-parameter causal LSTM that receives the same frozen decoded actions, input time deltas, packet order, and PCAP boundaries, but predicts complete raw packet fields, including output timing. The LSTM raises TCP-event distance from 12.66 to 43.52 percentage points, IAT TV from 0.49\% to 85.89\%, and flow-count error from 0.04\% to 631.22\%; valid-packet coverage falls from 100\% to 89.75\%. Under this shared action interface, the evaluated learned realizer does not match deterministic compilation. Appendix~\ref{app:learned_realizer} details its targets, paired results, and limitations.

\section{Conclusion}
This paper frames packet-trace representation learning as a codec-interface problem. PacketActions record packet behavior and timing without asking the neural codec to reproduce state-dependent packet fields. A packet-aligned VAE represents this interface, while a state-aware deterministic compiler realizes each decoded action under protocol state.
Our experiments show that this interface reduces the reconstruction failures observed for the evaluated raw-field baselines. \reprename{} does not reconstruct application-layer payload semantics, but it provides a direct path from generator-facing latents to timed packet traces realized under explicit flow and transport state.

\clearpage

\section*{AI Use Statement}
Generative AI assistance was used during manuscript preparation for language
editing, LaTeX organization, and consistency checks. The authors reviewed and
verified the resulting text, equations, citations, and experimental claims and
remain responsible for the submitted manuscript.

\section*{Ethics Statement}
This work uses publicly available network-trace datasets and does not involve
human subjects, user studies, or the collection of new personally identifying
information. The codec suppresses raw endpoint identifiers and payload bytes
and uses synthetic endpoint and payload realization. Privacy-relevant field
handling and its limitations are discussed in Appendix~\ref{app:privacy}.

\section*{Reproducibility Statement}
Public dataset sources, temporal splits, preprocessing,
training budgets, hardware, and evaluation commands are documented in
Appendices~\ref{app:setup} and~\ref{app:repro}; the main experimental artifacts
are generated from the shared scripts and recorded run manifests.

\bibliographystyle{tracecodec}
\bibliography{references}

\appendix

\section{Experimental Details}
\label{app:setup}

\subsection{Published baselines and source traces}

We compare \reprename against four published generic tabular models, each adapted to the shared raw packet-table interface described in Appendix~\ref{app:baselines}.

\textbf{TVAE}~\citep{Xu2019CTGAN} is the variational autoencoder baseline introduced alongside CTGAN for mixed-type tabular data. It learns a continuous row latent and reconstructs numerical and categorical columns directly.

\textbf{TabSyn-VAE}~\citep{Zhang2024TabSyn} is the VAE backbone used by TabSyn before latent-space diffusion. It tokenizes mixed-type rows and learns a Transformer-based latent representation; in our experiments, we use this VAE stage as a standalone codec rather than the full TabSyn generator.

\textbf{GOGGLE}~\citep{Liu2023GOGGLE} is a graph-based tabular generator that learns relational structure among columns and decodes rows through a learned graph model.

\textbf{TTVAE}~\citep{Wang2025TTVAE} is a Transformer-based variational autoencoder for tabular data. It replaces the MLP-style tabular backbone with a Transformer encoder--decoder.

The three source datasets contain public packet captures from distinct traffic regimes:
\begin{itemize}[leftmargin=1.5em]
    \item \textbf{CICIDS2017 Monday}~\citep{Sharafaldin2018CICIDS2017} is the benign Monday WorkingHours trace from the CICIDS2017 intrusion-detection dataset released by the Canadian Institute for Cybersecurity. We use it as an enterprise-style TCP-heavy setting with clear session structure.
    \item \textbf{MAWI \texttt{202004071400}}~\citep{Cho2000WIDERepository} is a public 15-minute backbone trace from the MAWI Working Group Traffic Archive. We use the samplepoint-F capture collected on 2020-04-07 14:00 as the higher-concurrency backbone setting.
    \item \textbf{VNAT}~\citep{Jorgensen2024VNAT} contains VPN and non-VPN captures from multiple network applications. We use it to test the codec on application traffic outside the enterprise and backbone settings above.
\end{itemize}

\subsection{Datasets and split policy}

All evaluation settings are built from raw packet traces rather than pre-aggregated flow summaries. For CICIDS2017 Monday and MAWI, we run a session-preserving temporal sharder, sort the resulting shard PCAPs by start time, and assign contiguous train/validation/test ranges over the ordered shard list. This keeps held-out shards temporally disjoint while preserving whole sessions inside each sample. The MAWI sharder produces 13,102 ordered shards; we use the first 3,000 and split them into 2,400/300/300 train, validation, and test shards.

For VNAT, we use 107/9/12 disjoint source captures for training, validation, and testing. After applying the shared packet-level scope, the three partitions contain 6.53M, 0.81M, and 1.27M PacketActions, respectively. Table~\ref{tab:eval_datasets} summarizes all three experimental splits.

\begin{table}[H]
\centering
\caption{Dataset statistics for the selected experimental splits after applying the shared packet-level scope. PCAPs are session-preserving shards for CICIDS2017 and MAWI and source captures for VNAT.}
\label{tab:eval_datasets}
\scriptsize
\setlength{\tabcolsep}{3.5pt}
\renewcommand{\arraystretch}{1.14}
\begin{tabular}{@{}l
                r
                rrr
                rrr
                rrr@{}}
\toprule
\textbf{Dataset} &
\textbf{PCAPs} &
\multicolumn{3}{c}{\textbf{Split PCAPs}} &
\multicolumn{3}{c}{\textbf{Tokens}} &
\multicolumn{3}{c}{\textbf{Protocol mix}} \\
\cmidrule(lr){3-5}
\cmidrule(lr){6-8}
\cmidrule(l){9-11}
&
&
\textbf{Train} &
\textbf{Val} &
\textbf{Test} &
\textbf{Train} &
\textbf{Val} &
\textbf{Test} &
\textbf{TCP} &
\textbf{UDP} &
\textbf{ICMP} \\
\midrule
CICIDS2017 Monday &
2,397 &
1,919 & 239 & 239 &
10.32M & 0.68M & 0.66M &
92.0\% & 8.0\% & 0.02\% \\
MAWI \texttt{202004071400} &
3,000 &
2,400 & 300 & 300 &
39.23M & 1.70M & 1.46M &
48.4\% & 33.0\% & 18.5\% \\
VNAT &
128 &
107 & 9 & 12 &
6.53M & 0.81M & 1.27M &
40.4\% & 59.6\% & 0.02\% \\
\bottomrule
\end{tabular}
\end{table}

\subsection{Shared preprocessing and decode scope}

All learned methods consume the same train, validation, and test partitions and are evaluated only after decoding their own outputs back to PCAP. The tokenizer, PacketAction schema, compiler, payload policy, and active-flow budget are fixed before training. Packets outside the shared packet-/transport-level decode scope are excluded consistently and recorded in the dataset manifests. Raw-field baselines use the same partitions but decode through the restricted renderer in Appendix~\ref{app:baselines}, so they do not inherit \reprename's state-aware compiler.

\paragraph{Ablation construction.}
The coarse-timing variant quantizes positive inter-packet deltas into eight bins that are uniform after the transform \(\log(1+\delta/\tau)\). Here, \(\tau\) is the median positive training delta, and the upper endpoint is the 99.9th percentile. The transport-cue ablation removes relative TCP sequence and acknowledgment progression and the advertised-window fields, while retaining the TCP option profile. All other PacketAction fields and the compiler are unchanged.

\subsection{Architecture and model capacity}
\label{app:architecture_capacity}

All neural codecs are trained from scratch on each dataset, without an external pretrained backbone. Saved model paths refer only to locally trained checkpoints. \reprename{} uses the same architecture on all three datasets. Its encoder embeds fields into 128-dimensional tokens and applies a six-layer, eight-head Transformer independently to each packet record. Each Transformer layer has feed-forward width 1536 and dropout 0.15. The resulting record representation is projected through hidden width 768 to the mean and log variance of a 128-dimensional Gaussian latent. Field-specific classification heads reconstruct the fixed-width PacketAction, and a separate regression head reconstructs timing. The deterministic compiler contains no trainable parameters.

\begin{table}[H]
\centering
\caption{Trainable parameter counts for the evaluated neural codecs. Counts are taken from the selected checkpoints. The compiler is deterministic and has no trainable parameters.}
\label{tab:model_capacity}
\small
\setlength{\tabcolsep}{8pt}
\renewcommand{\arraystretch}{1.08}
\begin{tabular}{@{}lrrr@{}}
\toprule
\textbf{Method} &
\textbf{CICIDS2017 Monday} &
\textbf{MAWI} &
\textbf{VNAT} \\
\midrule
TVAE        & 6,321,644  & 4,691,342   & 379,646 \\
TabSyn-VAE  & 1,627,481  & 70,159,385  & 177,716 \\
GOGGLE      & 14,791,699 & 631,578,835 & 1,743,814 \\
TTVAE       & 8,435,780  & 6,805,478   & 2,493,782 \\
\textbf{\reprename} & \textbf{11,494,401} & \textbf{11,494,401} & \textbf{11,494,401} \\
Compiler    & 0          & 0           & 0 \\
\bottomrule
\end{tabular}
\end{table}

Table~\ref{tab:model_capacity} reports exact trainable parameter counts from the evaluated checkpoints. The raw-field baseline counts vary across datasets because their learned categorical vocabularies depend on the observed packet fields. Model size therefore does not track decoded-PCAP fidelity: for example, on MAWI, TabSyn-VAE and GOGGLE are respectively 6.1 and 55.0 times larger than \reprename, while \reprename{} is larger than several baselines on VNAT. In the latent-interface experiment, the downstream flow-matching models contain 135,217,281 parameters for \reprename, 135,215,744 for TVAE, and 135,154,264 for TabSyn-VAE. Their maximum parameter-count difference is below 0.05\%.

\subsection{Runtime and resource use}
\label{app:runtime}

We report each reconstruction stage separately because neural inference is batched, whereas packet realization follows capture order. Training uses one NVIDIA RTX PRO 6000 GPU with bfloat16 and includes validation and checkpointing. Neural throughput is measured over the first 120,000 PacketActions from each test split with batch size 30,000, after one 30,000-packet warm-up batch. The input tensors are already resident on the GPU; checkpoint loading, host-to-device transfer, and disk I/O are excluded.

\begin{table*}[t]
\centering
\caption{Runtime and resource use across datasets. Training times are approximate. Neural throughput is measured on one NVIDIA RTX PRO 6000 GPU; compile/render throughput uses one Intel Xeon Platinum 8480C CPU process.}
\label{tab:runtime_performance}
\footnotesize
\setlength{\tabcolsep}{5pt}
\renewcommand{\arraystretch}{1.08}
\begin{tabular*}{\textwidth}{@{\extracolsep{\fill}}lrrr@{}}
\toprule
\textbf{Measurement} &
\textbf{CICIDS2017 Monday} &
\textbf{MAWI} &
\textbf{VNAT} \\
\midrule
Training time for 50k steps (h) & 13.0 & 12.2 & 11.4 \\
Neural encode (k packets/s) & 110.7 & 110.5 & 110.7 \\
Neural decode (M packets/s) & 2.84 & 2.82 & 2.84 \\
Stateful compile/render (k packets/s) & 1.48 & 1.40 & 1.75 \\
Peak GPU allocation (GiB) & 9.06 & 9.06 & 9.06 \\
\bottomrule
\end{tabular*}
\end{table*}

Neural encoding sustains 110.5--110.7k packets/s, and neural decoding reaches 2.82--2.84M packets/s. The ordered compile/render stage processes 1.40--1.75k packets/s in one CPU process and is the main measured cost in the current reconstruction implementation. This measurement includes action materialization, flow and transport-state updates, packet construction, and byte serialization, but excludes writing bytes to PCAP files and subsequent TShark analysis. It uses a 5,000-packet warm-up followed by six 20,000-packet measurement chunks. A 50,000-step training run takes approximately 11.4--13.0 hours and reaches a peak GPU allocation of 9.06 GiB.

The compiler operates online in capture order. Its mutable state is bounded by the configured number of active flow slots rather than the total trace length.

\section{Privacy Considerations}
\label{app:privacy}

This paper is not a formal privacy paper. It does not provide differential privacy, immunity to memorization, or jurisdiction-specific legal compliance. The privacy goal is specific: reduce direct disclosure of endpoint identifiers and application content while preserving the packet- and transport-level behavior needed by the codec.

\subsection{Identifier suppression}

\reprename keeps the variables needed for packet-/transport-level decode, such as timing, flow-slot identity, protocol, direction, control semantics, bounded acknowledgment advancement, receive-window structure, and payload-length behavior, but it does not learn raw endpoint identifiers or application payload bytes. In particular, the learned representation omits raw IP addresses, raw transport ports, MAC addresses, checksums, absolute TCP sequence and acknowledgment numbers, and payload content.

The practical implication is that the training representation is not a lightly disguised copy of the original packet headers. It is a structured behavioral representation whose purpose is to preserve packet- and transport-level semantics while suppressing directly identifying fields. Table~\ref{tab:privacy_scope} summarizes what is kept and what is omitted.

\begin{table}[t]
\centering
\caption{Privacy-relevant field treatment in \reprename. The codec keeps behavior needed for packet-/transport-level fidelity while suppressing direct identifiers and payload content.}
\label{tab:privacy_scope}
\footnotesize
\setlength{\tabcolsep}{4pt}
\renewcommand{\arraystretch}{1.10}
\begin{tabular}{p{0.30\linewidth} p{0.62\linewidth}}
\toprule
\textbf{Trace Element} & \textbf{Treatment in \reprename} \\
\midrule
Raw IP addresses, transport ports, MAC addresses &
Not part of the learned packet-action representation; decode re-instantiates synthetic endpoints from flow-slot identity rather than restoring original identifiers. \\
\addlinespace[2pt]
Application payload bytes &
Not learned. Decoded packets carry deterministic synthetic payload material by default, and truncated-capture output is also available when analysis-only artifacts are sufficient. \\
\addlinespace[2pt]
Absolute TCP sequence/ack numbers, checksums &
Not learned. They are reconstructed deterministically from relative state and packet rendering rules during decode. \\
\addlinespace[2pt]
Fidelity-relevant behavior: timing, direction, protocol, control events, payload length, bounded transport detail &
Retained, because these variables define the packet-/transport-level semantics the codec is intended to preserve. \\
\bottomrule
\end{tabular}
\end{table}

\subsection{Synthetic realization}

The decode path is synthetic rather than restorative. Flow templates are generated from flow-slot identity, protocol, IP family, generation index, and an optional salt; the resulting endpoints are therefore derived synthetic templates rather than original source and destination tuples. In the current implementation, IPv4 decode produces private-address-like endpoints, IPv6 decode produces unique-local-style endpoints, transport ports are regenerated, MAC addresses are synthetic, and TCP initial sequence numbers are re-instantiated from deterministic hashed state rather than copied from the source trace.

Payload handling follows the same principle. Payload bytes are not recovered from the training trace. Instead, the renderer emits deterministic synthetic payload material of the required length, and an analysis-oriented truncated-capture mode can further suppress payload bytes in the written PCAP while preserving wire length. The implementation also supports randomized per-run flow salts so that the same latent code can be decoded into different synthetic endpoint assignments across runs, improving unlinkability between decoded traces and any specific original address assignment.

\subsection{Scope}

These choices reduce direct identifier exposure and prevent literal replay of the original application payload. They are therefore meaningful privacy-aware design choices, and they are one reason we treat application-semantic replay as out of scope for the paper. However, they do not amount to a proof of privacy. Sensitive information can still survive indirectly through timing structure, payload-length patterns, rare protocol behavior, or distinctive interaction dynamics. A determined adversary could also study model memorization or nearest-neighbor leakage at the level of whole traces rather than individual identifiers.

For that reason, the correct interpretation of \reprename is not ``privacy guaranteed,'' but ``designed to suppress direct identifiers and payload content while preserving the semantics required for packet-/transport-level fidelity.'' Stronger privacy statements would require dedicated evaluation, such as membership-inference tests, nearest-neighbor disclosure audits, or formal mechanisms like differential privacy. Likewise, legal and institutional compliance remains a property of the full data-governance process, not of the model architecture alone.

\section{Raw-Field Baselines}
\label{app:baselines}

The main text compares \reprename against published generic tabular models adapted to raw-packet roundtrip. This appendix specifies that adaptation in enough detail to make the fairness assumptions explicit.

\subsection{Raw packet table}

Each packet is first converted into a mixed-type row that retains concrete header fields and timing rather than packet-action semantics. The design goal is not bit-identical reconstruction, but the strongest raw-field interface we can make compatible with protocol-constrained packet rendering without silently introducing stateful repair. The table therefore keeps fidelity-relevant concrete packet fields and omits obviously derived values that can be recomputed locally during rendering.

For the auxiliary raw-table diagnostics, we separate three pre-render quantities. Invalid-row rate counts decoded rows that cannot be interpreted as valid raw packet rows under the shared schema, for example because an inactive-token leaks into an active field or because the decoded endpoints are not valid IP literals for the declared IP family. On the remaining decoded rows that satisfy the shared contract, we report categorical-field accuracy over active categorical columns and numerical-field error over active numerical columns in raw field units. These are local diagnostics rather than decoded-PCAP metrics. Their purpose is to show whether a raw-field baseline has learned a non-trivial packet-row reconstruction map before rendering is judged, without conflating pre-render contract violations with later packet-tool failures.

The resulting schema is summarized in Table~\ref{tab:raw_packet_table_schema}. It is a direct per-packet projection of the protocol-neutral canonical packet contract used by the implementation rather than a bit-level dump or an engineered traffic-feature table. Source and destination IP addresses and transport ports are therefore retained as high-cardinality categorical columns, while \texttt{delta\_t}, fragment offset, payload length, TCP sequence and acknowledgment numbers, and TCP window remain numerical columns.

\begin{table}[t]
\centering
\caption{Raw packet-table schema used by tabular baselines.}
\label{tab:raw_packet_table_schema}
\scriptsize
\setlength{\tabcolsep}{3pt}
\renewcommand{\arraystretch}{0.98}
\begin{tabular}{@{}>{\raggedright\arraybackslash}p{0.28\linewidth}
                >{\raggedright\arraybackslash}p{0.64\linewidth}@{}}
\toprule
\textbf{Group} & \textbf{Columns and role} \\
\midrule
Timing &
\texttt{delta\_t}. First-class inter-packet timing variable. \\
\addlinespace[2pt]
Network / packet identity &
\texttt{ip\_version}, \texttt{proto}, \texttt{src\_ip}, \texttt{dst\_ip}, \texttt{ttl}, \texttt{dscp}, \texttt{ecn}, \texttt{ip\_flags}, \texttt{ip\_frag\_off}, \texttt{payload\_len}. Keep concrete family, endpoint, header, and payload-length fields needed for local packet rendering. \\
\addlinespace[2pt]
TCP block (TCP rows only) &
\texttt{tcp\_sport}, \texttt{tcp\_dport}, \texttt{tcp\_seq}, \texttt{tcp\_ack}, \texttt{tcp\_flags}, \texttt{tcp\_window}, \texttt{tcp\_opt\_profile}. Keep concrete transport values without introducing relative or state-derived semantics. \\
\addlinespace[2pt]
UDP block (UDP rows only) &
\texttt{udp\_sport}, \texttt{udp\_dport}. \\
\addlinespace[2pt]
ICMP block (ICMP rows only) &
\texttt{icmp\_type}, \texttt{icmp\_code}. \\
\bottomrule
\end{tabular}
\end{table}

Transport-specific columns are active only for rows whose \texttt{proto} value selects the matching block. Inactive categorical columns take a reserved ``not applicable'' token, while inactive numerical columns are set to zero and masked out of the local row-space diagnostics. The baseline therefore sees the same concrete packet fields that were present in the source packet and nothing more.

We exclude several tempting classes of columns. Locally derived quantities such as checksums, \texttt{ip\_total\_len}, \texttt{ip\_payload\_len}, \texttt{udp\_len}, and redundant presence bits are omitted because they can be recomputed once the retained columns are fixed. Derived cross-packet variables such as flow IDs, canonicalized direction, retransmission labels, and relative TCP progression are omitted because they would add sequence or state information to the raw-field baseline. Application payload bytes are excluded to match the packet-/transport-level scope of this study.

\subsection{Baseline methods}

We use four published tabular generative models as the primary raw-field baselines: TVAE \citep{Xu2019CTGAN}, TabSyn-VAE \citep{Zhang2024TabSyn}, GOGGLE \citep{Liu2023GOGGLE}, and TTVAE \citep{Wang2025TTVAE}. Each has a public implementation and a direct enough reconstruction path to support packet-wise encode--decode without redefining the task. In this paper they are evaluated narrowly as raw-field codecs rather than as unrestricted tabular synthesizers. All four consume the same train, validation, and test partitions used by \reprename{} and are evaluated after PCAP rendering. Their decoded raw rows use the restricted renderer described below, whereas \reprename{} realizes PacketActions through its state-aware compiler.

The selected baselines span distinct tabular modeling choices: TVAE and TTVAE are explicit variational codecs, TabSyn-VAE uses the VAE backbone of a recent latent-diffusion tabular model, and GOGGLE adds a learned-relational generator outside that VAE family. This selection evaluates the raw-field interface across both variational and relational architectures.

All raw-field baselines are trained with validation-selected checkpoints under fixed optimizer-step budgets, rather than with a fixed number of epochs. This choice makes the comparison stable under large raw-packet tables: an epoch over the CICIDS2017 Monday packet table contains millions of rows, so epoch counts are a dataset-size-dependent unit of computation. The validation objective is the model's raw-table reconstruction objective on the validation split, and the decoded test outputs are always produced from the best validation checkpoint. The step budgets and validation intervals are specified in Appendix~\ref{app:repro}; the run manifests record the final step, best step, stopping reason, validation objective, and full training configuration.

\paragraph{Latent-interface controls.}
The latent-flow experiment in the main text uses only raw-field baselines whose latent code can be decoded without access to the original input row. TVAE satisfies this criterion directly: the decoder maps a row latent vector to the transformed raw packet table. TabSyn-VAE also satisfies it after flattening its non-CLS field-token latents; the flattened vector can be generated by the same latent-flow model and reshaped before the TabSyn decoder and field heads recover a raw row.

TTVAE is not used as a latent-interface control because its decoder is not a function of the latent vector alone. In the adapted implementation, the encoder first maps the transformed row to a Transformer encoder output \(\mathbf{h}_{\mathrm{enc}}\), then projects that same output to \(\mu\) and \(\log\sigma^2\). The decoder maps \(z\) to decoder tokens but reconstructs through a Transformer decoder that cross-attends to \(\mathbf{h}_{\mathrm{enc}}\). Thus reconstruction has the form \(D(z,\mathbf{h}_{\mathrm{enc}}(x))\), not \(D(z)\). A downstream generator would therefore have to keep the input-derived encoder memory, generate a second high-dimensional memory sequence, or replace it with an arbitrary placeholder. Each choice changes the latent decode contract rather than testing whether a standalone raw-field latent sequence is useful.

GOGGLE is also excluded from the latent-flow control set. Although its model has a learned row-level latent, the validation-selected GOGGLE checkpoint produces no decoded rows satisfying the shared raw-packet table contract on CICIDS2017 Monday under the non-repair policy, as shown in Table~\ref{tab:raw_table_diagnostics}. Running a downstream generator on that latent space would therefore mostly reproduce the same pre-render contract failure rather than test latent sequence modelability. We keep GOGGLE in the main decoded-PCAP comparison, where that failure mode is part of the raw-field baseline evidence.

\subsection{Rendering policy}

To keep the comparison fair, baseline outputs are rendered back into packets under a restricted deterministic policy. Decoded rows are split back along the original reference PCAP boundaries before rendering, matching the evaluation convention used for \reprename. Local recomputation is allowed only for quantities that are straightforward consequences of a single decoded packet instance, such as checksums or directly derived length fields. In practice, packet libraries may materialize these packet-local derived values during serialization; we treat that as ordinary rendering rather than as heuristic repair. Sequence-level or flow-state repair is not allowed. In particular, the baseline renderer may not impute missing active fields, infer hidden endpoints or transport state, repair decoded sequence or acknowledgment progression, or borrow \reprename's packet-action compiler logic. Rows that already violate the raw-table contract are counted as invalid decoded rows and are not repaired before rendering. This restriction is what makes the comparison meaningful: if raw-field baselines require stateful repair to become parseable, the repair has become part of the representation problem.

A protocol state machine is deterministic only once its input event is well specified. Independently decoded raw fields may fail to specify a unique legal event under the current flow state: flags, sequence progress, acknowledgment progress, endpoint orientation, timing, and payload length can disagree. A stateful renderer must then choose a precedence rule. If it trusts the raw fields, it may emit inconsistent packets; if it trusts protocol state and overwrites raw fields, it is no longer evaluating raw-field reconstruction; if it chooses case by case, it has become heuristic repair. Such a precedence rule is an implicit codec/compiler contract. The non-repair policy therefore isolates the raw-field interface rather than forbidding a stronger method: a raw baseline augmented with a full stateful compiler has moved toward the explicit interface design used by \reprename.

\subsection{Raw-table diagnostics}

Table~\ref{tab:raw_table_diagnostics} reports the pre-render raw-table diagnostics for the published baselines. These rows are deliberately not mixed into the main decoded-PCAP table: they are useful for explaining whether a raw-field model fails before rendering, but they are undefined for \reprename and the Oracle path because those methods do not decode through raw packet-table rows.

\begin{table}[t]
\centering
\caption{Pre-render validity diagnostics for raw-field baselines.}
\label{tab:raw_table_diagnostics}
\scriptsize
\setlength{\tabcolsep}{3.0pt}
\renewcommand{\arraystretch}{0.98}
\begin{tabular}{@{}l l r r r@{}}
\toprule
\textbf{Dataset} &
\textbf{Method} &
\textbf{Inv. row (\%) \(\downarrow\)} &
\textbf{Cat acc. (\%) \(\uparrow\)} &
\textbf{Num err. \(\downarrow\)} \\
\midrule
CICIDS2017 Monday &
TVAE &
37.46 &
71.68 &
\(4.12{\times}10^{8}\) \\
&
TabSyn-VAE &
2.39 &
84.74 &
\(2.79{\times}10^{7}\) \\
&
GOGGLE &
100.00 &
-- &
-- \\
&
TTVAE &
37.28 &
65.99 &
\(2.57{\times}10^{8}\) \\
\midrule
MAWI \texttt{202004071400} &
TVAE &
14.22 &
80.68 &
\(1.64{\times}10^{8}\) \\
&
TabSyn-VAE &
28.06 &
87.36 &
\(1.46{\times}10^{7}\) \\
&
GOGGLE &
100.00 &
-- &
-- \\
&
TTVAE &
61.97 &
67.17 &
\(5.26{\times}10^{7}\) \\
\bottomrule
\end{tabular}
\end{table}

\subsection{Oracle roundtrip}

We additionally run one no-learning oracle control:
\[
\text{PCAP} \rightarrow \text{raw packet table} \rightarrow \widehat{\text{PCAP}}.
\]
This is not a competing baseline. It simply checks whether the shared raw table and restricted renderer can already close a near-exact roundtrip when given ground-truth rows directly. We report the check on CICIDS2017 Monday and MAWI.

Its role is interpretive. If this oracle roundtrip is near-exact while learned raw-field baselines still decode poorly, the failure is not primarily in the single-packet contract or the renderer. It appears after learning, when a generic raw-field codec has to carry cross-packet behavior through its bottleneck without the state variables used by \reprename.

\begin{table*}[t]
\centering
\caption{Raw-table oracle roundtrip under the restricted renderer.}
\label{tab:raw_table_oracle_roundtrip}
\scriptsize
\setlength{\tabcolsep}{3.0pt}
\renewcommand{\arraystretch}{0.98}
\begin{tabular}{@{}l r r r r r r@{}}
\toprule
\textbf{Dataset} &
\textbf{TCP event dist. (pp) \(\downarrow\)} &
\textbf{Count (\%) \(\downarrow\)} &
\textbf{Proto (\%) \(\downarrow\)} &
\textbf{IAT TV (\%) \(\downarrow\)} &
\textbf{Flow cnt. (\%) \(\downarrow\)} &
\textbf{Flow dur. TV (\%) \(\downarrow\)} \\
\midrule
CICIDS2017 Monday &
0.002 &
0.42 &
0.42 &
0.13 &
0.09 &
0.64 \\
MAWI &
0.063 &
\(<0.001\) &
\(<0.001\) &
\(<0.001\) &
\(<0.001\) &
\(<0.001\) \\
\bottomrule
\end{tabular}
\end{table*}

\section{Metric Definitions}
\label{app:metrics}

This appendix defines the metrics reported in the main tables and figures. Percent columns multiply the underlying fraction or distance by 100. ``pp'' denotes percentage points. All decoded-PCAP metrics are computed after packet rendering, not on intermediate token sequences.

\paragraph{Shard-aligned decoded-PCAP evaluation.}
Unless otherwise stated, a decoded test split is compared against the matching reference shards \(\{x_s\}_{s=1}^S\). Packet analyzers run on each reference/decoded shard pair and the resulting summaries are aggregated. This avoids mixing independent capture units and keeps sessionization, TCP analysis, and packet-count accounting aligned across methods.

\paragraph{TCP event dist.}
The main-table TCP event distance compares the rate profile of structured TCP diagnostic events. Let \(\mathcal{A}\) contain retransmission, fast retransmission, spurious retransmission, lost segment, ACKed lost segment, duplicate ACK, out-of-order segment, and zero-window event. If \(c_a(x)\) and \(c_a(\hat{x})\) are the counts of event \(a\) in the reference and decoded PCAPs, and \(N_{\mathrm{tcp}}(x)\), \(N_{\mathrm{tcp}}(\hat{x})\) are their TCP packet counts, the table reports
\[
\sum_{a \in \mathcal{A}}
\left|
\frac{c_a(\hat{x})}{\max(N_{\mathrm{tcp}}(\hat{x}), 1)}
-
\frac{c_a(x)}{\max(N_{\mathrm{tcp}}(x), 1)}
\right|
\]
in percentage points. This is a profile-fidelity metric: real captures may contain TCP diagnostic events, so the target is the reference profile rather than zero events.

\paragraph{Count.}
If \(N\) and \(\hat{N}\) are the reference and decoded packet counts, the Count column reports
\[
\frac{|\hat{N} - N|}{\max(N, 1)}.
\]
This penalizes decoded traces that silently drop or invent packets. The latent-interface appendix labels the same quantity as Count err.

\paragraph{Proto.}
Let \(p\) and \(\hat{p}\) be protocol-ratio vectors over TCP, UDP, ICMP, and other packets. The Proto column reports total variation,
\[
\mathrm{TV}(p,\hat{p})=\frac{1}{2}\sum_i |p_i-\hat{p}_i|.
\]

\paragraph{IAT TV.}
Inter-arrival times are histogrammed using fixed millisecond bins shared by all methods. If \(h_{\mathrm{iat}}\) and \(\hat{h}_{\mathrm{iat}}\) are the normalized reference and decoded histograms, the IAT TV column reports \(\mathrm{TV}(h_{\mathrm{iat}},\hat{h}_{\mathrm{iat}})\).

\paragraph{Flow cnt.}
Flows are measured as bidirectional transport sessions under the same sessionization rule used for duration analysis. If \(M\) and \(\hat{M}\) are the reference and decoded session counts, the Flow cnt. column reports
\[
\frac{|\hat{M} - M|}{\max(M, 1)}.
\]
The value can exceed 100\% when a method fragments traffic into many more sessions than the reference.

\paragraph{Flow dur. TV.}
Each bidirectional five-tuple is split into transport sessions before duration measurement. TCP sessions end at RST/FIN boundaries or after a 60-second idle gap; UDP, ICMP, and other sessions use the same idle split. Session durations are histogrammed using fixed millisecond bins, and Flow dur. TV reports total variation between the reference and decoded duration histograms.

\paragraph{TCP transition distance.}
Figure~\ref{fig:tcp_transition_state_consistency} compares event-to-event TCP trajectories within bidirectional sessions. Each trace is converted into a row-normalized transition matrix over coarse TCP event labels. The scalar transition distance reported in the text is the weighted row-wise total variation between the decoded and reference matrices, with row weights taken from the reference transition mass.

\paragraph{Flow-switch and run-length diagnostics.}
Figure~\ref{fig:active_flow_interleaving} measures whether decoded traces preserve multi-flow interleaving. The packet-to-packet flow-switch rate is the fraction of adjacent decoded packets whose bidirectional session label changes. Same-flow run length is the length of a maximal consecutive packet run belonging to the same bidirectional session. Session count is the same sessionization count used by Flow cnt.

\paragraph{Action field acc.}
The ablation-table Action field acc. column is the mean PacketAction field accuracy for \reprename on the held-out split. It averages exact-match accuracy over decoded PacketAction fields under the active masks of the schema, so inactive protocol-specific fields do not contribute. This is an intrinsic codec diagnostic; decoded-PCAP metrics remain the primary fidelity evidence.

\paragraph{KL rate.}
The ablation-table KL rate reports the test posterior KL of the codec encoder to the unit-Gaussian prior,
\[
\mathbb{E}\left[
\sum_{k=1}^{d}
\frac{1}{2}\left(\mu_k^2+\exp(\log\sigma_k^2)-1-\log\sigma_k^2\right)
\right],
\]
in nats per packet over latent dimension \(d\). We report the raw posterior KL without free-nats clamping. The metric is an intrinsic latent-regularity diagnostic: removing KL can leave reconstruction metrics nearly unchanged while moving the posterior much farther from the unit-Gaussian prior. We do not train a separate downstream generator for each ablation.

\paragraph{Usable suffixes.}
In the latent-interface experiment, each case asks a downstream generator to produce a 1024-token suffix from a matched 1024-token prefix. Usable suffixes is the fraction of decoded cases whose rendered PCAP contains exactly the requested suffix packet count. It is a case-level measure of whether the sampled latent suffix remains on a complete decode surface.

\paragraph{Valid pkt coverage.}
Valid packet coverage is the decoded valid packet mass divided by the requested suffix packet mass, summarized over latent-interface cases. It gives partial credit when a suffix decodes to some valid packets but loses or invalidates the rest.

\paragraph{Coverage-adjusted TV.}
Latent-interface distributional metrics compare decoded suffixes with their matched held-out real suffixes. Let \(P_b\) and \(Q_b\) be the normalized reference and valid-decoded histograms. Let \(N\) be the reference packet count, \(M\) the decoded packet count before validity filtering, and \(V\) the valid decoded packet count. We define
\[
\operatorname{CA\mbox{-}TV}(P,Q)
=
\frac{1}{2}
\left[
\sum_b \left|P_b-\frac{V}{N}Q_b\right|
+ \frac{(M-V)_+}{N}
+ \frac{|N-M|}{N}
\right],
\]
with \(Q_b=0\) when \(V=0\). The first term measures distribution mismatch at the surviving mass; the remaining terms charge invalid packets and packet-count mismatch. When \(M=V=N\), the metric reduces to ordinary TV. The CA IAT TV, CA pkt-size TV, CA flow-pkts TV, and CA sess-dur TV columns apply this accounting to inter-arrival-time, packet-size, packets-per-flow, and session-duration histograms, respectively.

\paragraph{Raw-table diagnostics.}
The auxiliary raw-table table reports pre-render diagnostics for raw-field baselines only. Inv. row is the fraction of decoded rows that violate the shared raw-packet contract before packet rendering. Cat acc. is mean exact-match accuracy over active categorical raw-table columns on rows that survive the invalid-row filter. Num err. is mean absolute error over active numerical raw-table columns in raw field units. These diagnostics explain pre-render raw-table behavior and are not mixed into the main decoded-PCAP comparison.

\section{Extended Results}
\label{app:extended}

This appendix reports a paired comparison of deterministic and learned realization, followed by additional results on MAWI.

\subsection{Replacing deterministic compilation with a learned realizer}
\label{app:learned_realizer}

We compare deterministic and learned realization in a paired \(2 \times 2\) experiment. For each dataset, the action source is either the ground-truth PacketAction sequence or one reconstruction sampled from the trained codec and then frozen. Each sequence is realized by either the deterministic compiler or a learned causal model. The paired paths therefore receive identical actions, timing, packet order, and PCAP boundaries. Because each decoded source is a single stochastic reconstruction, compiler values in this table need not match values from a different reconstruction run in Table~\ref{tab:eval_main_results}; only within-source comparisons are paired.

The learned replacement is a one-layer causal LSTM with hidden width 768 and 4,272,965 trainable parameters. It receives all 36 PacketAction fields and \(\delta_t\), then predicts 22 complete raw packet-table fields. Its training objective covers all source-PCAP targets, including endpoint literals, ports, and absolute sequence and acknowledgment values. PacketActions deliberately omit some of these source-specific values, so they are not uniquely determined by the input action sequence. We therefore evaluate the structure of the rendered PCAP rather than exact source-field recovery and treat the experiment as a comparison with this particular full-field learned replacement. We train a separate realizer for each dataset on 256-packet segments for three epochs with batch size 32 and AdamW at learning rate 0.001. Epoch 3 is selected by validation loss. The causal hidden state crosses inference chunks and resets only at PCAP boundaries.

The learned path does not call the \flowtable, TCP state machines, compiler constraints, or sequence repair. Its packet-local serializer only selects protocol-specific fields, accumulates predicted timing, computes lengths and checksums, and writes PCAP bytes. Thus, it can materialize a predicted row but cannot repair flow identity or transport state across packets.

\begin{table*}[t]
\centering
\caption{Paired realization comparison across three datasets. Within each action-source pair, the compiler and LSTM receive identical actions, timing, packet order, and PCAP boundaries. Bold marks the better result within a pair; ties are unbolded. TCP distance is in percentage points; all other metrics are percentages. Lower is better except for valid coverage and IP parseability.}
\label{tab:learned_realizer_2x2}
\scriptsize
\setlength{\tabcolsep}{2.6pt}
\renewcommand{\arraystretch}{1.02}
\resizebox{\textwidth}{!}{%
\begin{tabular}{@{}lllrrrrrrrr@{}}
\toprule
\textbf{Dataset} &
\textbf{Action source} &
\textbf{Realizer} &
\textbf{TCP event dist.} &
\textbf{Count} &
\textbf{Proto} &
\textbf{IAT TV} &
\textbf{Flow cnt.} &
\textbf{Flow dur. TV} &
\textbf{Valid cov.} &
\textbf{IP parse} \\
\midrule
CICIDS2017 & Ground truth & Compiler &
\textbf{2.17} &
\textbf{0.00} &
\textbf{0.00} &
\textbf{0.04} &
\textbf{0.00} &
\textbf{0.01} &
\textbf{100.00} &
100.00 \\
CICIDS2017 & Ground truth & Causal LSTM &
43.04 &
2.02 &
0.76 &
86.64 &
694.35 &
77.30 &
97.98 &
100.00 \\
CICIDS2017 & Decoded & Compiler &
\textbf{12.66} &
\textbf{0.00} &
\textbf{0.00} &
\textbf{0.49} &
\textbf{0.04} &
\textbf{9.13} &
\textbf{100.00} &
100.00 \\
CICIDS2017 & Decoded & Causal LSTM &
43.52 &
10.25 &
3.95 &
85.89 &
631.22 &
77.61 &
89.75 &
100.00 \\
\addlinespace[2pt]
MAWI & Ground truth & Compiler &
\textbf{1.55} & \textbf{0.00} & \textbf{0.00} & \textbf{0.01} & \textbf{0.00} & \textbf{0.00} & \textbf{100.00} & 100.00 \\
MAWI & Ground truth & Causal LSTM &
74.19 & 0.04 & 0.02 & 85.89 & 51.20 & 14.36 & 99.96 & 100.00 \\
MAWI & Decoded & Compiler &
\textbf{2.32} & \textbf{0.00} & \textbf{0.00} & \textbf{0.39} & \textbf{0.07} & \textbf{0.25} & 100.00 & 100.00 \\
MAWI & Decoded & Causal LSTM &
54.86 & 0.00 & 22.51 & 79.71 & 59.38 & 16.40 & 100.00 & 100.00 \\
\addlinespace[2pt]
VNAT & Ground truth & Compiler &
\textbf{0.00} & \textbf{0.00} & \textbf{0.00} & \textbf{0.02} & \textbf{0.00} & \textbf{0.00} & \textbf{100.00} & 100.00 \\
VNAT & Ground truth & Causal LSTM &
20.11 & 28.51 & 22.69 & 92.68 & 28{,}628.51 & 69.46 & 71.49 & 100.00 \\
VNAT & Decoded & Compiler &
\textbf{0.14} & \textbf{0.00} & \textbf{0.01} & \textbf{1.09} & \textbf{7.46} & \textbf{6.32} & \textbf{100.00} & 100.00 \\
VNAT & Decoded & Causal LSTM &
0.67 & 100.00 & 28.03 & 98.59 & 97.33 & 76.28 & 0.002 & 100.00 \\
\bottomrule
\end{tabular}
}
\end{table*}

Table~\ref{tab:learned_realizer_2x2} compares the two realization paths while holding each action source fixed. On CICIDS2017, replacing the compiler with the LSTM raises TCP-event distance from 12.66 to 43.52 percentage points and flow-count error from 0.04\% to 631.22\% for the same decoded actions. On MAWI, the replacement raises TCP-event distance from 2.32 to 54.86 percentage points and IAT TV from 0.39\% to 79.71\%. On VNAT, only 0.002\% of the LSTM output from decoded actions satisfies the shared packet contract, compared with 100\% for compiler output. Its low TCP-event distance is therefore not evidence of faithful realization because almost no predicted rows survive the packet contract.

All compiler outputs have 100\% valid coverage. Every LSTM output remains 100\% IP-parseable, but its failures differ by dataset, including flow fragmentation, timing distortion, and invalid packet records. Parseability alone therefore does not preserve sessions or transport behavior. This single-seed control does not show that neural realizers can never learn protocol state. It shows that this causal learned replacement does not match deterministic compilation under the same action interface.

\subsection{Additional MAWI results}

The results below use the MAWI \texttt{202004071400} samplepoint-F trace and the temporal split described in Appendix~\ref{app:setup}: 2,400 train shards, 300 validation shards, and 300 test shards, corresponding to 39.23M/1.70M/1.46M PacketAction tokens after the shared packet-level scope.

\begin{figure*}[t]
\centering
\includegraphics[width=0.94\textwidth]{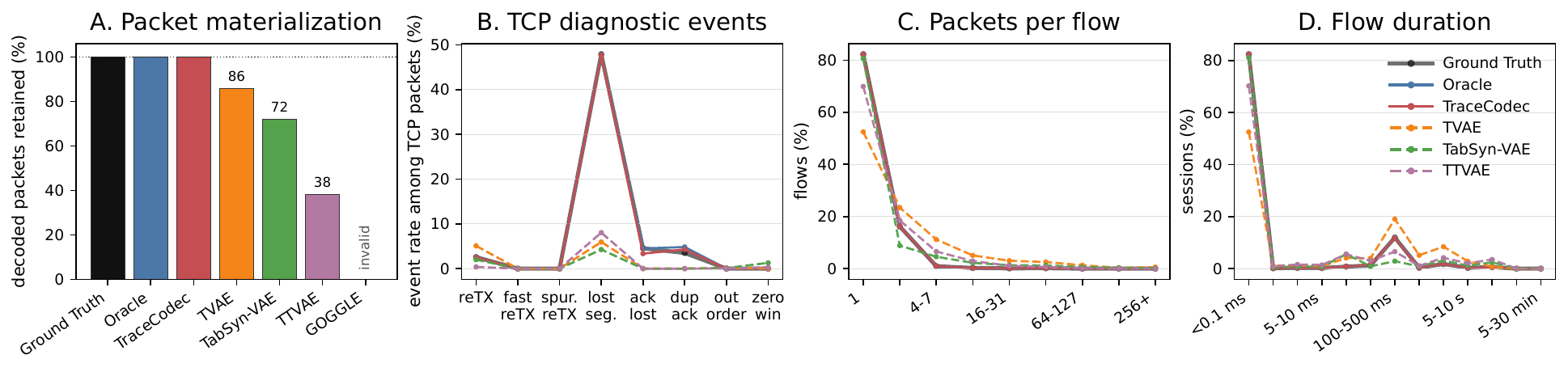}
\caption{Protocol-state diagnostics on MAWI, using the panels from Figure~\ref{fig:decoded_pcap_claim_panels}.}
\label{fig:mawi_decoded_pcap_claim_panels}
\end{figure*}

\begin{figure*}[t]
\centering
\includegraphics[width=0.70\textwidth]{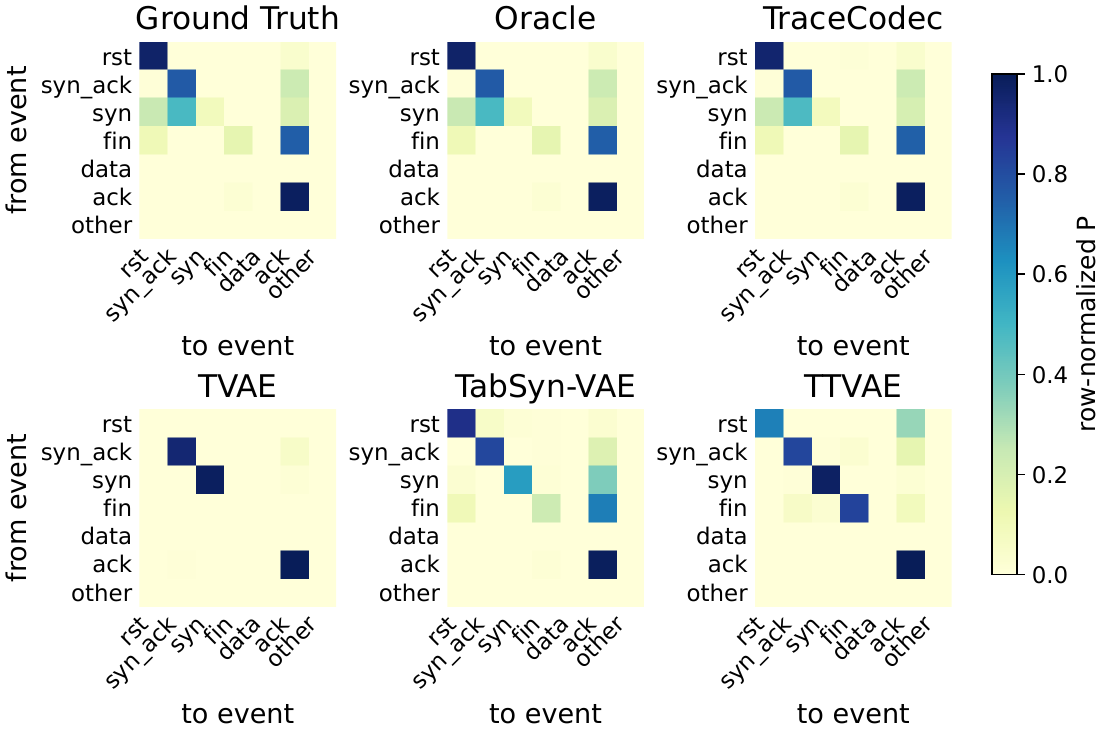}
\caption{TCP transition consistency on MAWI. Each heatmap is a row-normalized transition matrix over coarse TCP event labels within bidirectional sessions, matching Figure~\ref{fig:tcp_transition_state_consistency}.}
\label{fig:mawi_tcp_transition_state_consistency}
\end{figure*}

\begin{figure*}[t]
\centering
\includegraphics[width=\textwidth]{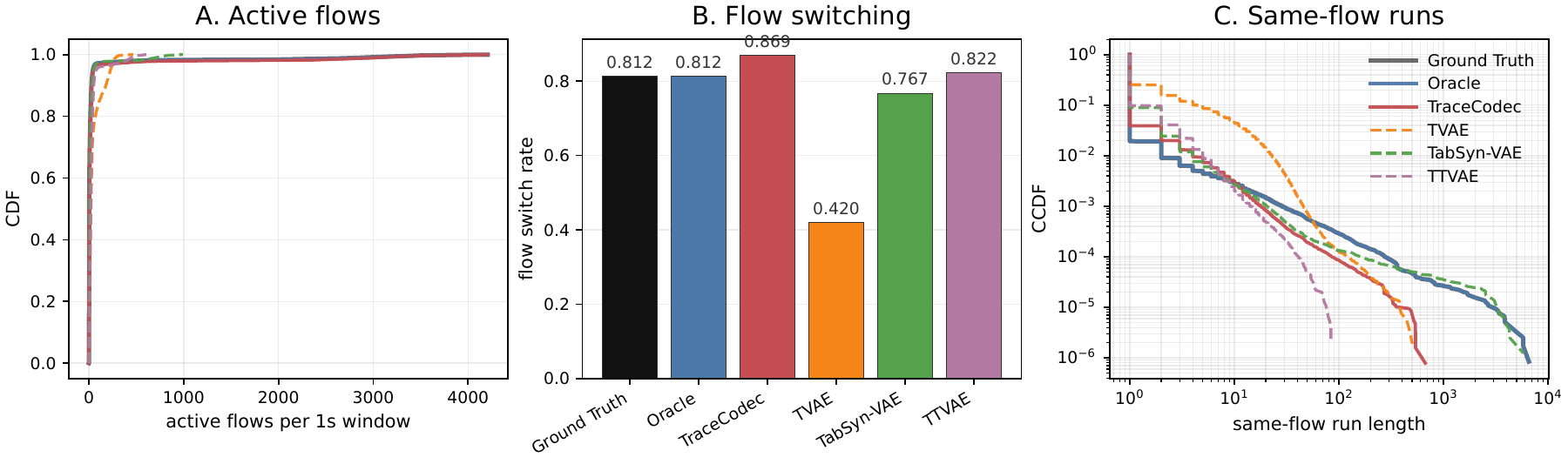}
\caption{Active-flow and multi-flow interleaving on MAWI, using the diagnostics from Figure~\ref{fig:active_flow_interleaving}.}
\label{fig:mawi_active_flow_interleaving}
\end{figure*}

Figures~\ref{fig:mawi_decoded_pcap_claim_panels}--\ref{fig:mawi_active_flow_interleaving} apply the decoded-PCAP and structural diagnostics from the main text to the MAWI held-out split. \reprename preserves packet materialization, TCP-event structure, and multi-flow progression more consistently than the raw-field baselines, which either lose packet mass or fragment transport and session structure after rendering.

\begin{table*}[t]
\centering
\caption{MAWI ablation results for \reprename. Boxes mark targeted metrics; KL rate is the encoder posterior KL to the unit-Gaussian prior, summed over latent dimensions and averaged per packet.}
\label{tab:eval_ablations_mawi}
\scriptsize
\setlength{\tabcolsep}{2.6pt}
\renewcommand{\arraystretch}{1.02}
\newcommand{\mawiabmark}[1]{\begingroup\setlength{\fboxsep}{1.0pt}\setlength{\fboxrule}{0.45pt}\fbox{\strut #1}\endgroup}
\begin{tabular*}{\textwidth}{@{\extracolsep{\fill}}l r r r r r r@{}}
\toprule
\textbf{Variant} &
\shortstack{\textbf{Action field}\\\textbf{acc. (\%) \(\uparrow\)}} &
\shortstack{\textbf{TCP event}\\\textbf{dist. (pp) \(\downarrow\)}} &
\shortstack{\textbf{IAT TV}\\\textbf{(\%) \(\downarrow\)}} &
\shortstack{\textbf{Flow count}\\\textbf{err. (\%) \(\downarrow\)}} &
\shortstack{\textbf{Flow duration}\\\textbf{TV (\%) \(\downarrow\)}} &
\shortstack{\textbf{KL rate}\\\textbf{(nats/pkt)}} \\
\midrule
Full \reprename &
98.75 &
2.31 &
0.38 &
0.07 &
0.25 &
249.38 \\
\textsc{w/o} KL &
99.50 &
3.33 &
0.34 &
0.04 &
0.17 &
\mawiabmark{5829.68} \\
\textsc{w/o} fine timing &
99.05 &
2.81 &
\mawiabmark{8.36} &
0.07 &
0.41 &
141.66 \\
\textsc{w/o} TCP cues &
99.67 &
\mawiabmark{115.29} &
0.32 &
0.04 &
0.19 &
71.40 \\
\bottomrule
\end{tabular*}
\end{table*}

Table~\ref{tab:eval_ablations_mawi} shows the same interface roles as the main ablation study. Coarsening timing raises IAT TV from 0.38\% to 8.36\%, while flow-count error remains at 0.07\%. Removing TCP progression and window cues leaves action accuracy high but increases TCP-event distance from 2.31 to 115.29 percentage points. Removing KL keeps the decoded reconstruction metrics close but raises the test posterior KL rate from 249.38 to 5829.68 nats per packet. As on CICIDS2017 Monday, reconstruction fidelity and latent regularity are distinct properties.

\subsection{Latent-interface experiment}

The main text reports only a compact latent-interface check. This appendix gives the details needed to interpret that table. Each PCAP shard is first encoded into a packet-ordered latent sequence. Fixed prefix/suffix windows are then extracted with a 1024-token prefix, a 1024-token generated suffix, and stride 512. The setting is long enough to require continuation of timing and multi-flow interaction structure, but still small enough to train the same downstream model across representations.

The latent generator is a Transformer velocity model trained with conditional flow matching. It attends over the concatenation of prefix latents and noisy suffix latents, receives a scalar flow-time embedding, and predicts the suffix velocity. The reported latent objective is validation flow-matching MSE under fixed validation noise and time draws. For decoded evidence, sampled suffix latents are decoded back through the corresponding representation and compared against real held-out suffix PCAPs materialized from the same continuation windows, using the same decoded-PCAP diagnostics reported elsewhere in the paper. This evaluation asks whether a generator can stay on a usable decode surface while matching the actual continuation trace, rather than looking good only relative to a representation-specific oracle decode.

For \reprename{}, each sequence element concatenates the normalized codec latent with a one-dimensional transformed time delta; the downstream model generates both components. The raw-field controls use their standalone latent codes, whose decoders reconstruct timing as a raw field.

For \reprename, sampled latents are lowered through the fixed packet-action decoder and compiler. Suffix decoding starts from an empty runtime and replays the decoded within-window prefix, without emitting it, to rebuild flow and transport state before emitting the sampled suffix. For raw-field controls, sampled TVAE and TabSyn-VAE latents are decoded through their own raw-table decoders and then rendered under the same non-repair policy used in the main comparison. TTVAE and GOGGLE are not included in this latent-interface control set for the reasons described in Appendix~\ref{app:baselines}: TTVAE's decoder requires input-derived encoder memory in addition to \(z\), and GOGGLE already fails the raw-row contract before rendering on CICIDS2017 Monday.

\begin{table*}[t]
\centering
\caption{Latent-interface check on MAWI. CA: coverage-adjusted total variation.}
\label{tab:latent_interface_mawi}
\scriptsize
\setlength{\tabcolsep}{4.0pt}
\renewcommand{\arraystretch}{0.98}
\resizebox{\textwidth}{!}{%
\begin{tabular}{@{}l r r r r r r@{}}
\toprule
\textbf{Representation} &
\textbf{Usable suffixes (\%) \(\uparrow\)} &
\textbf{Valid pkt coverage (\%) \(\uparrow\)} &
\textbf{CA IAT TV (\%) \(\downarrow\)} &
\textbf{CA pkt-size TV (\%) \(\downarrow\)} &
\textbf{CA flow-pkts TV (\%) \(\downarrow\)} &
\textbf{CA sess-dur TV (\%) \(\downarrow\)} \\
\midrule
\textbf{\reprename} &
{\bfseries 98.44} &
{\bfseries 100.00} &
{\bfseries 5.38} &
{\bfseries 1.42} &
{\bfseries 4.33} &
{\bfseries 5.12} \\
TVAE &
7.03 &
85.69 &
40.22 &
16.60 &
39.21 &
41.04 \\
TabSyn-VAE &
10.94 &
62.30 &
43.33 &
40.38 &
45.22 &
46.95 \\
\bottomrule
\end{tabular}
}
\end{table*}

Table~\ref{tab:latent_interface_mawi} repeats the same continuation protocol on MAWI. \reprename again provides the most reliable decode surface, with nearly all suffixes usable, full median valid-packet coverage, and substantially lower coverage-adjusted timing, packet-size, flow-size, and session-duration distances than the raw-field latent controls. TVAE and TabSyn-VAE can produce parseable packets, but their sampled suffixes lose packet mass on many windows; the coverage-adjusted diagnostics therefore expose larger continuation errors than ordinary distribution distances alone.

The main ablation table uses the intrinsic posterior KL rate defined in Appendix~\ref{app:metrics} to measure latent regularity. The separate flow-matching experiment evaluates whether the resulting generator-facing representation supports traffic continuation.

\section{Reproducibility Details}
\label{app:repro}

This appendix records the public entry points and run metadata needed to reproduce the main result tables. The implementation writes JSON manifests at every dataset-build, training, reconstruction, and metric-aggregation stage, so each reported row can be traced back to the corresponding run output.

\paragraph{Software environment.}
All experiments are run from the repository root in a Python 3.10+ environment created for the project. A typical setup is:
\begin{verbatim}
conda create -n trace_codec python=3.10
conda activate trace_codec
python -m pip install --upgrade pip
python -m pip install -e .
python -m pip install jupyter ipykernel nbclient nbformat
python -m ipykernel install --user --name trace_codec \
  --display-name "trace_codec"
\end{verbatim}
The package dependencies are declared in \texttt{pyproject.toml}; the extra notebook packages are only needed to execute the paper notebooks. Decoded-PCAP diagnostics use Wireshark's command-line tools, so \texttt{tshark} must also be available on \texttt{PATH}. For example, on Ubuntu:
\begin{verbatim}
sudo apt-get update
sudo apt-get install -y tshark
\end{verbatim}
The paper runs used CUDA-enabled PyTorch on NVIDIA RTX PRO 6000 Blackwell Server Edition GPUs with 96GB memory. The scripts expose \texttt{--device cuda} and can be restricted to a particular GPU with \texttt{CUDA\_VISIBLE\_DEVICES}. Runtime depends on the selected stages; the training budgets and evaluation caps are specified below and recorded in run manifests.

\paragraph{Notebook entry points.}
The main CICIDS2017 Monday workflow is launched from \texttt{trace\_codec\_experiment.ipynb}; the MAWI and VNAT workflows use \texttt{trace\_codec\_experiment\_mawi.ipynb} and \texttt{trace\_codec\_experiment\_vnat.ipynb}, respectively. After activating the environment above, the notebooks can be run interactively with the \texttt{trace\_codec} kernel or executed unattended:
\begin{verbatim}
python -u scripts/execute_notebook_with_checkpoints.py \
  trace_codec_experiment.ipynb \
  --output outputs/trace_codec_experiment.executed.ipynb \
  --kernel-name trace_codec \
  --timeout -1 \
  2>&1 | tee outputs/trace_codec_experiment.nbconvert.log
\end{verbatim}
Replace the notebook path and output log with the MAWI or VNAT names to reproduce those runs. The notebooks call the scripts listed below and skip expensive stages when their sentinel artifacts already exist.

\paragraph{Shared split and dataset construction.}
Within each dataset, all learned methods consume the same train, validation, and test partitions described in Appendix~\ref{app:setup}. For the raw-field baselines, the shared packet table is built once and then reused by TVAE and TTVAE; TabSyn-VAE and GOGGLE derive their prepared numerical/categorical arrays from that same table:
\begin{verbatim}
bash scripts/run_packet_table_baseline_experiments.sh \
  --dataset-key cic_ids2017_monday \
  --split-root dataset/trace_codec_experiments/cic_ids2017_monday/split \
  --device cuda \
  --baselines tvae,tabsyn_vae,goggle,ttvae
\end{verbatim}
The command can be restricted to individual stages with \texttt{--phases build}, \texttt{--phases train}, or \texttt{--phases reconstruct}. Reconstructions optionally include \texttt{tshark} expert-analysis diagnostics with \texttt{--tshark-expert}.

\paragraph{Raw-field baseline training protocol.}
Raw-field baselines use validation-selected checkpoints under fixed optimizer-step budgets. This avoids treating one pass over a multi-million-row packet table as the primitive unit of comparison. The default budgets are TVAE: 100k optimizer steps, TTVAE: 100k steps, GOGGLE: 100k steps, and TabSyn-VAE: 15k steps. TVAE, TTVAE, and GOGGLE are evaluated every 5k steps; TabSyn-VAE is evaluated every 1k steps. Training stops at the step budget or after eight consecutive non-improving validation events, with \texttt{min\_delta=0}. The selected checkpoint minimizes the validation raw-table reconstruction objective, and all held-out reconstructions are decoded from that checkpoint.

The corresponding script options are:
\begin{verbatim}
--max-steps tvae=100000,tabsyn_vae=15000,goggle=100000,ttvae=100000
--eval-interval-steps tvae=5000,tabsyn_vae=1000,goggle=5000,ttvae=5000
--patience-evals 8
--min-delta 0.0
\end{verbatim}
Each training manifest records the full configuration, \texttt{global\_step}, \texttt{best\_step}, \texttt{stop\_reason}, \texttt{best\_val\_objective}, and the final held-out reconstruction diagnostics. The accompanying \texttt{history.jsonl} file records every validation event, including the training-window averages since the previous validation event. For GOGGLE, learned-graph thresholding is also keyed to optimizer step, and reconstruction materializes the graph at the selected checkpoint step.

\paragraph{\reprename training and artifact mapping.}
\reprename dataset construction, training, decode, and evaluation are launched from \texttt{trace\_codec\_experiment.ipynb}, which calls the repository scripts under the same Python environment. We train for at most 50,000 optimizer steps with AdamW, effective batch size 30,000, learning rate \(10^{-4}\), and a cosine schedule with 2,000 warmup steps. Training uses bfloat16, seed 42, KL weight \(5\times10^{-4}\) with a 16-epoch warmup, 0.01 free nats per latent dimension, and a timing-loss weight of 4.0 after clipping time deltas at 10 seconds. We validate every 1,000 steps and export the checkpoint with the lowest validation loss.

Generated outputs are written under \path{outputs/trace_codec_experiments/<dataset>/<variant>/}; run metadata and training curves are written under \path{runs/trace_codec_experiments/<dataset>/<variant>/}. The main comparison rows are summarized from the reconstructed test PCAP reports, while intrinsic codec diagnostics and latent checks are written under the corresponding variant output directories.

The posterior KL-rate diagnostic in the ablation table is computed directly from the exported codec checkpoint and test split:
\begin{verbatim}
OUT=outputs/trace_codec_experiments/cic_ids2017_monday/full
python scripts/compute_trace_codec_latent_rate.py \
  --model-dir runs/trace_codec_experiments/cic_ids2017_monday/full/model \
  --split test \
  --batch-size 30000 \
  --device cuda \
  --output "$OUT/latent_rate_test.json"
\end{verbatim}
The script reads the dataset path and dataset-variant contract from the run artifacts, then reports raw posterior KL without free-nats clamping.

\paragraph{Latent-interface training protocol.}
The latent-interface experiment is launched by the same notebooks through \texttt{scripts/run\_latent\_sequence\_experiments.sh}. For each representation, the script first exports shard-aligned latent sequences for the shared train/validation/test split. A \reprename{} sequence element is \([\mu_t; y_t]\), where \(\mu_t\) is the normalized codec latent and \(y_t=\log(1+\delta_t/\tau)\) is the explicit time coordinate using the codec's training-set time scale \(\tau\). The flow model predicts both components with objective \(\mathcal{L}_{\mu}+0.04\mathcal{L}_{y}\); generated timing is decoded from the predicted \(y_t\), not from the VAE timing head. The script then trains a conditional latent-flow model on fixed prefix/suffix windows with stride 512. The paper-facing CICIDS2017 Monday setting uses a 1024-token prefix and a 1024-token generated suffix. Unless otherwise stated, the flow model is a 12-layer Transformer with hidden width 768, 12 attention heads, MLP ratio 4, dropout 0.05, logit-normal flow-time sampling, bfloat16 autocast, no activation checkpointing, and AdamW with learning rate \(2\times 10^{-4}\), weight decay \(10^{-2}\), gradient clipping at 1.0, 1000 warmup steps, cosine decay to a 0.2 learning-rate ratio, and EMA decay 0.9999. The completed run uses batch size 80, 32k optimizer steps, and validation/checkpointing every 500 steps. Checkpoints are selected by the lower validation flow-matching MSE between raw and EMA weights; manifests record the selected source, step, and full configuration. Evaluation samples suffixes with a 400-step Heun solver, reports latent metrics over up to 512 windows, decodes up to 128 suffix cases into PCAPs, and compares those predicted suffix PCAPs with real held-out suffix PCAPs materialized from the same continuation windows.

The corresponding command shape is:
\begin{verbatim}
bash scripts/run_latent_sequence_experiments.sh \
  --dataset-key cic_ids2017_monday \
  --split-root dataset/trace_codec_experiments/cic_ids2017_monday/split \
  --device cuda \
  --representations trace_codec,tvae,tabsyn_vae \
  --prefix-length 1024 \
  --horizon 1024 \
  --stride 512 \
  --train-batch-size 80 \
  --train-max-steps 32000 \
  --train-eval-steps 500 \
  --train-lr 2e-4 \
  --lr-scheduler cosine \
  --warmup-steps 1000 \
  --min-lr-ratio 0.2 \
  --ema-decay 0.9999 \
  --d-model 768 \
  --num-layers 12 \
  --num-heads 12 \
  --dropout 0.05 \
  --time-loss-weight 0.04 \
  --sample-steps 400 \
  --sample-solver heun \
  --max-metric-windows 512 \
  --max-decoded-cases 128 \
  --tshark-expert
\end{verbatim}

\end{document}